\documentclass[pre,aps,amsmath,amssymb]{revtex4}
\usepackage{mathrsfs}
\usepackage{amssymb}
\usepackage{graphicx}
\usepackage{notes2bib}

\usepackage{amsmath}
\usepackage[export]{adjustbox}
\usepackage{graphicx}
\usepackage{dcolumn}
\usepackage{bm}
\newcommand{\bra}[1]{\left(#1\right)}
\newcommand{\brb}[1]{\left[#1\right]}
\newcommand{\brc}[1]{\left<#1\right>}

\newcommand{\fw}{0.8}
\newcommand{\hfw}{0.4}

\newcommand{\be}{\begin{equation}}
\newcommand{\ee}{\end{equation}}
\newcommand{\bea}{\begin{eqnarray}}
\newcommand{\eea}{\end{eqnarray}}

\newcommand{\e}{\text{e}}

\newcommand{\edges}{\mathcal{E}}
\newcommand{\fBethe}{f_{\textrm{Bethe}}}
\newcommand{\betaSG}{\beta_\textrm{SG}}
\newcommand{\betaR}{\beta_\textrm{R}}
\newcommand{\eps}{\epsilon}

\newcommand{\cin}{c_ {\textrm{in}}}
\newcommand{\cout}{c_ {\textrm{out}}}

\newcommand{\omegain}{\omega_{\text{in}}}
\newcommand{\omegaout}{\omega_{\text{out}}}

\newcommand{\argmax}{\arg\!\max}
\newcommand{\myvspace}[1]{}

\newcommand{\betanish}{\beta_{\textrm{Nishimori}}}

\begin{document}
\title{Scalable detection of statistically significant communities and hierarchies, using message-passing for modularity}
\author{Pan Zhang and Cristopher Moore}
\affiliation{
Santa Fe Institute, Santa Fe, New Mexico 87501, USA}
\begin{abstract}
Modularity is a popular measure of community structure.  However, maximizing the modularity can lead to 
many competing partitions, with almost the same modularity, that are poorly correlated with each other.  
It can also produce illusory ``communities'' in random graphs where none exist.  We address 
this problem by using the modularity as a Hamiltonian at finite temperature, and using an efficient Belief Propagation algorithm to obtain the consensus of many partitions with high modularity, rather than looking for a single partition that maximizes it.  
We show analytically and numerically that the proposed algorithm works all the way down to the 
detectability transition in networks generated by the stochastic block model. It also 
performs well on real-world networks, revealing large communities in 
some networks where previous work has claimed no communities exist.  
Finally we show that by applying our algorithm recursively, 
subdividing communities until no statistically-significant subcommunities can be found,
we can detect hierarchical structure in real-world networks more efficiently than previous methods.

\smallskip
\emph{Significance}:  
Most work on community detection does not address the issue of statistical significance, and many algorithms are prone to overfitting. We address this using tools from statistical physics.  Rather than trying to find the partition of a network that maximizes the modularity, our approach seeks the consensus of many high-modularity partitions.  We do this with a scalable message-passing algorithm, derived by treating the modularity as a Hamiltonian and applying the cavity method.  We show analytically that our algorithm succeeds all the way down to the detectability transition in the stochastic block model; it also performs well on real-world networks.  It also provides a principled method for determining the number of groups, or hierarchies of communities and subcommunities.
\end{abstract}
\maketitle

Community detection, or node clustering, is a key problem in network science, computer science, sociology, and biology.  It aims to partition the nodes in a network into groups such that there are many edges connecting nodes within the same group, and comparatively few edges connecting nodes in different groups. 

Many methods have been proposed for this problem.  These include spectral clustering, where we classify nodes according to the eigenvectors of a linear operator such as the adjacency matrix, random walk matrix, graph Laplacian, or other linear operators~\cite{Luxburg07atutorial,PhysRevE.74.036104,Krzakala24122013}; statistical inference, where we fit the network with a generative model such as the stochastic block model~\cite{hastings,decelle-etal-pre,decelle-etal-prl,karrer-newman}; and a wide variety of other methods, e.g.~\cite{clauset2004finding,blondel2008fast,rosvall2008maps}.  See~\cite{Santo201075} for a review.

We focus here on a popular measure of the quality of a partition, the modularity (e.g.~\cite{newman-girvan,newman2004fast,clauset2004finding,duch2005community}).  A partition into $q$ groups is a set of labels $\{t\}$, where $t_i \in \{1,\ldots,q\}$ is the group to which node $i$ belongs.  The modularity of a partition $\{t\}$ of a network with $n$ nodes and $m$ edges 
is defined as follows, 
\begin{equation}
\label{eq:mod}
	Q(\{t\})=\frac{1}{m}\bra{\sum_{\brc{ij}\in \edges}\delta_{t_it_j}-\sum_{\brc{ij}}\frac{d_id_j}{2m}\delta_{t_it_j}} \, .
\end{equation}
Here 
$\edges$ is the set of edges, the degree $d_i$ is the number of neighbors node $i$ has, and $\delta$ is the Kronecker delta.  Thus $Q$ is proportional to the number of edges within communities, minus the expected number of such edges if the graph were randomly rewired while keeping the degrees fixed; that is, the expectation in a null model where $i$ and $j$ are connected with probability $d_i d_j/2m$.

However, maximizing over all possible partitions often gives a large modularity even in random graphs with no community structure~\cite{guimera2004modularity,reichardt2006statistical,zdeborova2010conjecture,sulc-zdeborova}.  
Thus maximizing the modularity can lead to overfitting, where the ``optimal'' partition simply reflects random noise.  
Even in real-world networks, the modularity often exhibits a large amount of degeneracy, with multiple local optima that are poorly correlated with each other, and are not robust to small perturbations~\cite{good2010performance}.  

\begin{figure}
   \centering
    \includegraphics[width=\hfw\columnwidth]{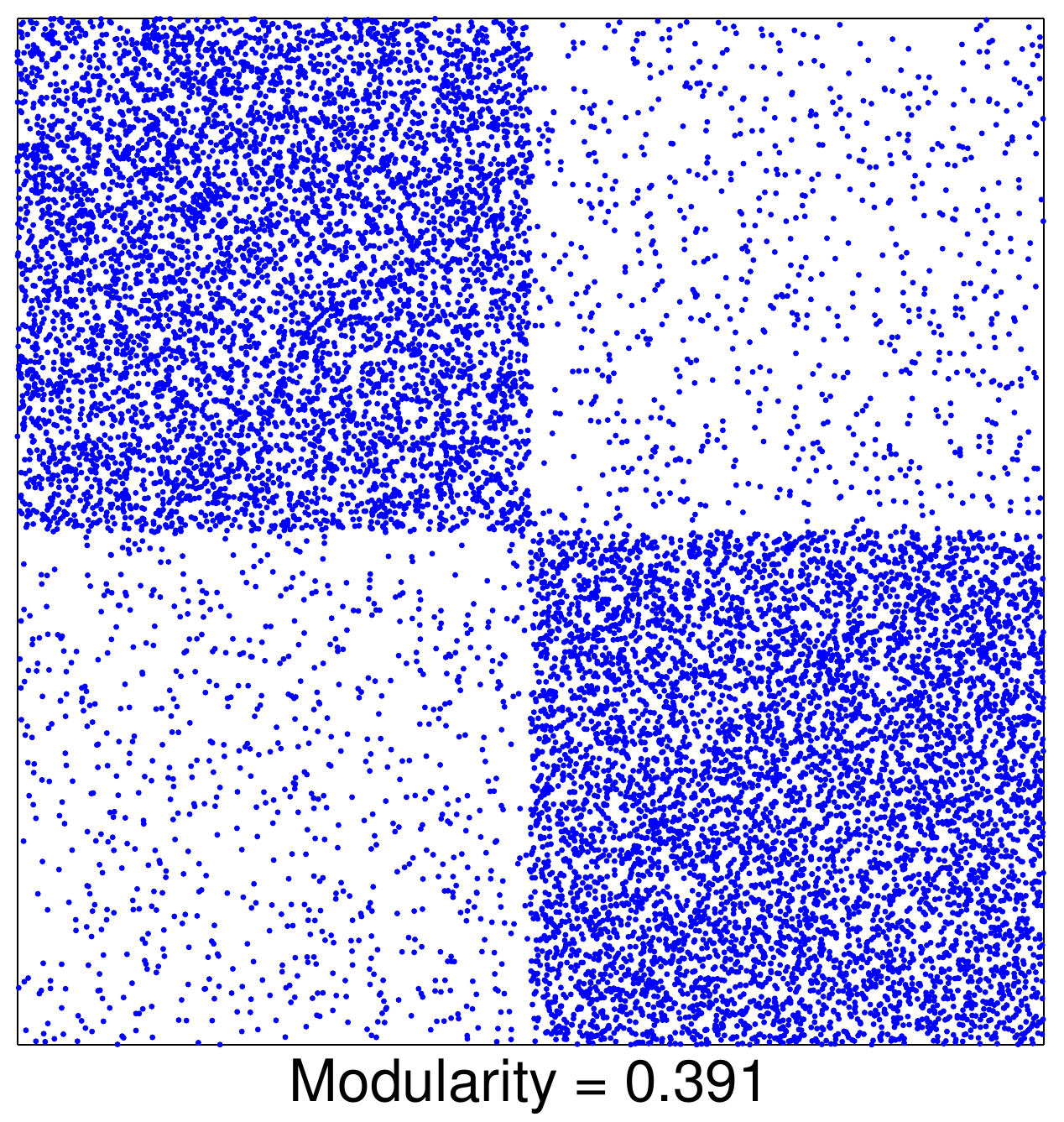} 
	\includegraphics[width=\hfw\columnwidth]{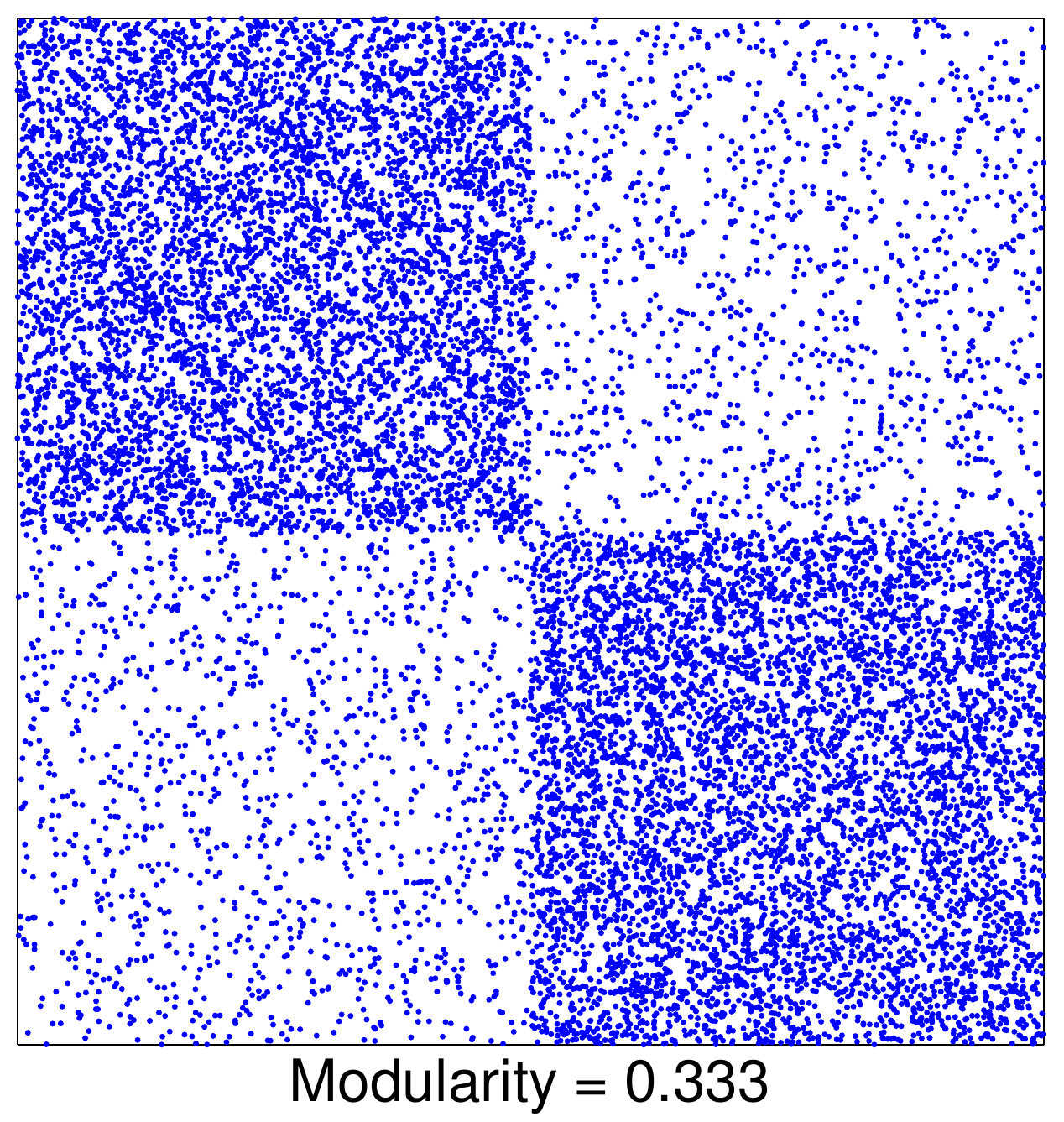} 
	\caption{The adjacency matrices of two networks, partitioned to show possible community structure.  Each blue point is an edge.  The network on the left is an ER graph, with no real community structure; however, a search by simulated annealing finds a partition with modularity $0.391$.  The network on the right has true communities, and is generated by the stochastic block model, but the true partition has modularity just $0.333$.  Thus illusory communities in random graphs can have higher modularity than true communities in structured graphs.  Both networks have size $n=5000$ and a Poisson degree distribution with mean $c=3$; the network on the right has $\cout/\cin=0.2$, in the easily-detectable regime of the stochastic block model.
	\label{fig:two_A}}
\end{figure}

Thus we need to add some notion of statistical significance to our algorithms.  One approach is hypothesis testing, comparing various measures of community structure to the distribution we would see in a null model such as Erd\H{o}s-R\'enyi (ER) graphs~\cite{Lancichinetti-pre-2010,Lancichinetti-plosone,wilson-testing}.  However, even when communities really exist, the modularity of the true partition is often no higher than that of random graphs.  In Fig.~\ref{fig:two_A}, we show partitions of two networks with the same size and degree distribution: an ER graph (left), and a graph generated by the stochastic block model (right), in the detectable regime where it is easy to find a partition correlated with the true one~\cite{decelle-etal-pre,decelle-etal-prl}.  The true partition of the network on the right has a smaller modularity than the partition found for the random graph on the left.  We can find a partition with higher modularity (and lower accuracy) on the right using e.g. simulated annealing, but then the modularities we obtain for the two networks are similar.  Thus the usual approach of null distributions and $p$-values for hypothesis testing does not appear to work.

We propose to solve this problem with the tools of statistical physics.  Like~\cite{reichardt2006statistical}, we treat the modularity as the Hamiltonian of a spin system.  We define the energy of a partition $\{t\}$ as
$E(\{t\}) = -m Q(\{t\})$, 
and introduce a Gibbs distribution as a function of inverse temperature $\beta$, 
$P(\{t\}) \propto \e^{-\beta E(\{t\})}$.  
Rather than maximizing the modularity by searching for the ground state of this system, we focus on its Gibbs distribution at a finite temperature, looking for many high-modularity partitions rather than a single one.  In analogy with previous work on the stochastic block model~\cite{decelle-etal-pre,decelle-etal-prl}, we define a partition $\{\hat{t}\}$ by computing the marginals of the Gibbs distribution, and assigning each node to its most-likely community.  Specifically, if $\psi^i_t$ is the marginal probability that $i$ belongs to group $t$, then $\hat{t}_i = \argmax_{t}\psi_{t}^i$, 
breaking ties randomly if more than one $t$ achieves the maximum.  
We call $\{\hat{t}\}$ the \emph{retrieval partition}, and call its modularity $Q(\{ \hat{t} \})$ the \emph{retrieval modularity}.  We claim that $\{ \hat{t} \}$ is a far better measure of significant community structure than the maximum-modularity partition.  In the language of statistics, the maximum marginal prediction is better than the maximum a posteriori prediction (e.g.~\cite{iba1999}).  More informally, the consensus of many good solutions is better than the ``best'' single one~\cite{clauset2008hierarchical,Lancichinetti-consensus}. 

We give an efficient Belief Propagation (BP) algorithm to approximate these marginals, which is derived from the cavity method of statistical physics.  This algorithm is highly scalable; each iteration takes linear time on sparse networks if the number of groups is fixed, 
and it converges rapidly in most cases.  It is optimal in the sense that for synthetic graphs generated by the stochastic block model, it works all the way down to the detectability transition.  
It provides a principled way to choose the number of communities, unlike other algorithms that tend to overfit.  
Finally, by applying this algorithm recursively, subdividing communities until no statistically significant subcommunities exist, we can uncover hierarchical structure.

We validate our approach with experiments on real and synthetic networks.  In particular, we find significant large communities in some large networks where previous work claimed there were none.  We also compare our algorithm with several others, finding that it obtains more accurate results, both in terms of determining the number of communities and matching their ground truth structure.


\section{Results}\label{sec:results}
\subsection{Results on the Stochastic Block Model}\label{sec:results_SBM}

Also called the planted partition model, the stochastic block model (SBM) is a popular ensemble of networks with community structure.  There are $q$ groups of nodes, and each node $i$ has a group label $t^*_i \in \{1, \ldots, q\}$; thus $\{t^*\}$ is the true, or planted, partition.  Edges are generated independently according to a $q \times q$ matrix $p$, by connecting each pair of nodes $\brc{ij}$ with probability $p_{t^*_i,t^*_j}$. Here for simplicity we discuss the commonly studied case where the $q$ groups have equal size and where $p$ has only two distinct entries, $p_{rs}=\cin / n$ if $r=s$ and $\cout / n$ if $r \ne s$.  We use $\eps=\cout/\cin$ to denote the ratio between these two entries.  In the assortative case, $\cin > \cout$ and $\eps < 1$.  When $\eps$ is small, the community structure is strong; when $\eps=1$, the network becomes an ER graph.

For a given average degree $c=(\cin + (q-1) \cout)/q$, there is a so-called detectability phase transition~\cite{decelle-etal-prl,decelle-etal-pre}, at a critical value 
\begin{equation}
\label{eq:eps}
\eps^* = \frac{\sqrt c -1}{\sqrt c -1 +q} \, . 
\end{equation}
For $\eps < \eps^*$, BP can label the nodes with high accuracy; for $\eps > \eps^*$, neither BP nor any other algorithm can label the nodes better than chance, and indeed no algorithm can distinguish the network from an ER graph with high probability.  This transition was recently established rigorously in the case $q=2$~\cite{mossel2012stochastic,massoulie2013community,mossel2013proof}.  

For larger numbers of groups, the situation is more complicated.  For $q \le 4$, in the assortative case, this detectability transition coincides with the Kesten-Stigum bound~\cite{kesten_stigum_1,kesten_stigum_2}.  For $q \ge 5$ the Kesten-Stigum bound marks a conjectured transition to a ``hard but detectable'' phase where community detection is still possible but takes exponential time, while the detectability transition is at a larger value of $\eps$; that is, the thresholds for reconstruction and robust reconstruction become different.  Our claim is that our algorithm succeeds down to the Kesten-Stigum bound, i.e., throughout the detectable regime for $q \le 4$ and the easily detectable regime for $q \ge 5$.

\begin{figure}
   \centering
    \includegraphics[width=\hfw\columnwidth]{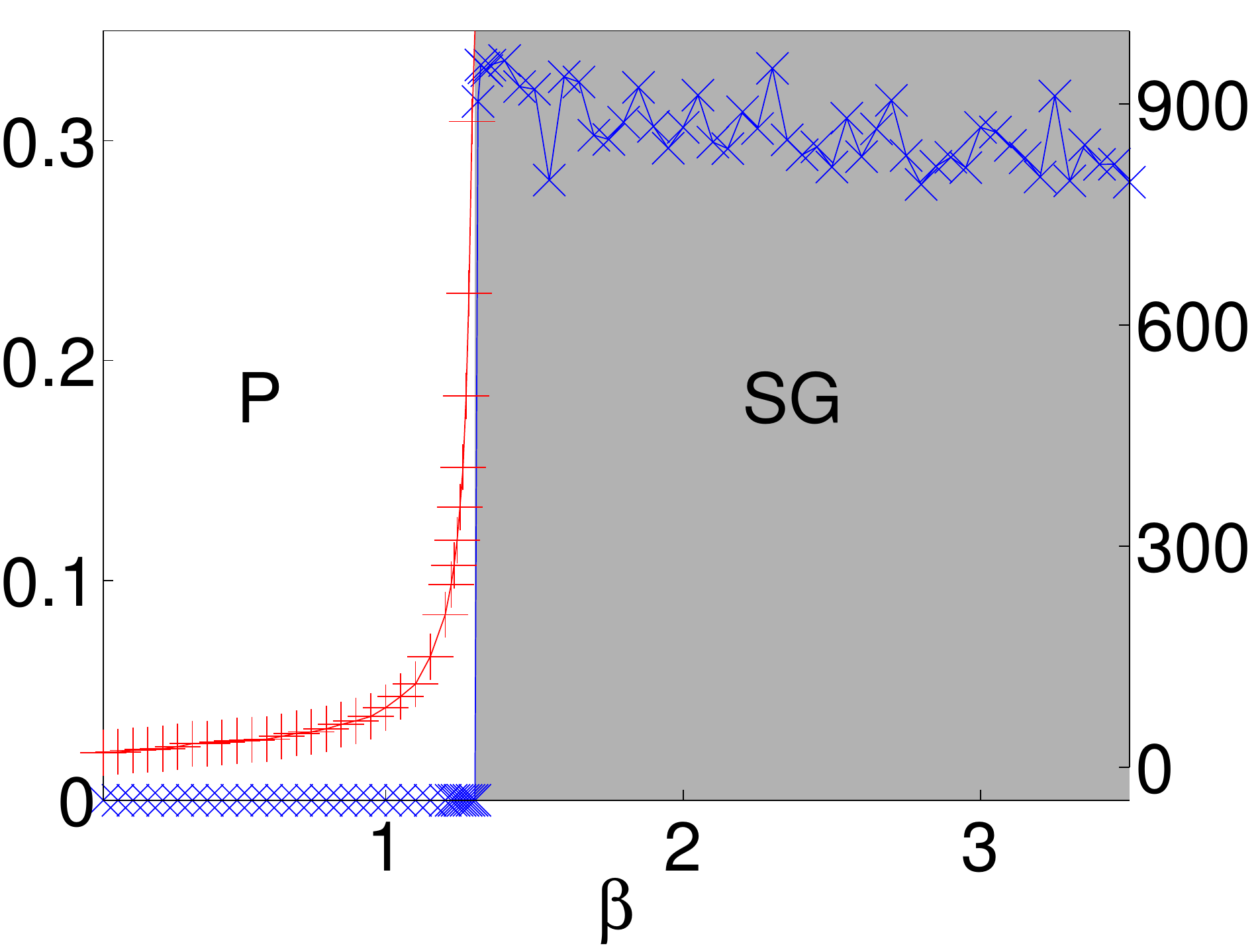} 
	\includegraphics[width=\hfw\columnwidth]{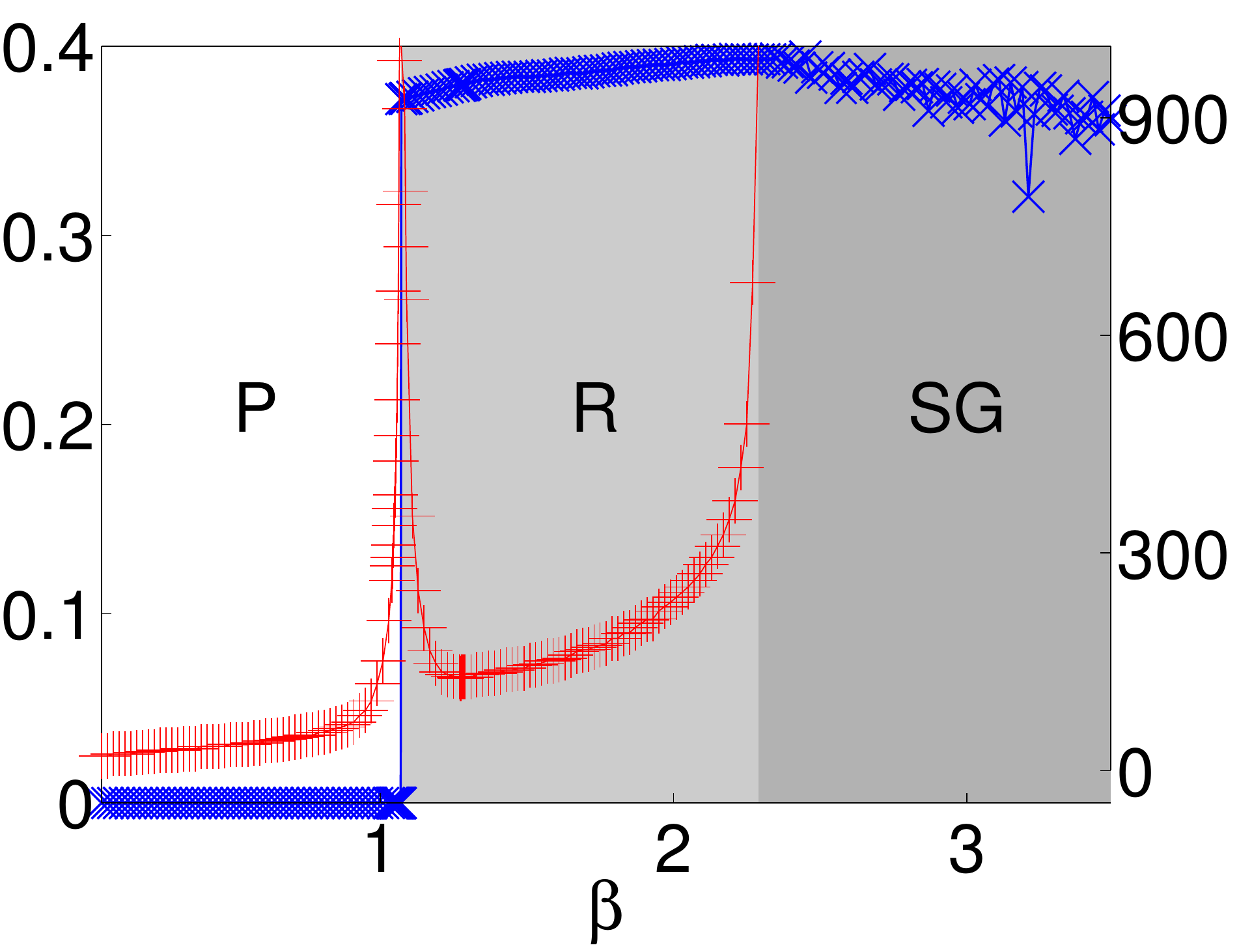} 
\caption{Retrieval modularity (blue $\times$, left y-axis) and BP convergence time (red $+$, right y-axis) of an ER random graph (left) and a network generated by the stochastic block model in the detectable regime (right). Both networks have $n=1000$ and average degree $c=3$, and the network on the right has $\eps=0.2$.  In both cases we ran BP with $q=2$ groups.  In the ER graph, which has no community structure, there are two phases, paramagnetic (P) and spin glass (SG), with a transition at $\beta^* = 1.317$.  In the SBM network, there is an additional retrieval phase (R) between $\betaR = 1.072$ and $\betaSG = 2.27$ where BP finds a retrieval state with high modularity, indicating statistically significant community structure.  
	\label{fig:two_phases}}
\end{figure}

In Fig.~\ref{fig:two_phases} we compare the behavior of our BP algorithm on ER graphs and a network generated by the SBM in the detectable regime.  Both graphs have the same size and average degree $c=3$.  For the ER graph (left) there are just two phases, separated by a transition at $\beta^* = 1.317$: the \emph{paramagnetic phase} where BP converges to a factorized fixed point where every node is equally likely to be in every group, and the \emph{spin glass phase} where replica symmetry is broken, and BP fails to converge.   The convergence time diverges at the transition.  Note that in the spin glass phase, the retrieval modularity returned by BP fluctuates wildly as BP jumps from one local optimum to another, and has little meaning.  In any case BP assumes replica symmetry, which is incorrect in this phase.

In contrast, the SBM network in Fig.~\ref{fig:two_phases} (right) has strong community structure.  In addition to the paramagnetic and spin glass phases, there is now a \emph{retrieval phase} in a range of $\beta$, where BP finds a retrieval state describing statistically significant community structure.  The retrieval modularity jumps sharply at $\betaR = 1.072$ when we first enter this phase, and then increases gently to $0.393$ as $\beta$ increases; for comparison, the modularity of the planted partition is $M_{\text{hidden}}(\eps)=1/(1+\eps)-1/2=0.33$.  When we enter the spin glass phase at $\betaSG = 2.27$, the retrieval modularity fluctuates as in the ER graph.  The convergence time diverges at both phase transitions.  

We can compute two of these transition points analytically by analyzing the linear stability of the factorized fixed point (see Methods).  Stability against random perturbations gives
\begin{equation}
\label{eq:beta-star}
\beta^*(q,c) = \log \left( \frac{q}{\sqrt{c} -1} + 1 \right) \, ,
\end{equation}
and stability against correlated perturbations gives
\begin{equation}
\label{eq:beta_eps}
	\betaR(q,c,\eps) = \log\bra{ \frac{q (1+(q-1) \eps)}{c(1-\eps)-(1+(q-1)\eps)}+1} \, .
\end{equation}
These cross at the Kesten-Stigum bound, where $\eps = \eps^*$.  We do not currently have an analytic expression for $\betaSG$. 

\begin{figure}
   \centering
    \includegraphics[width=\hfw\columnwidth]{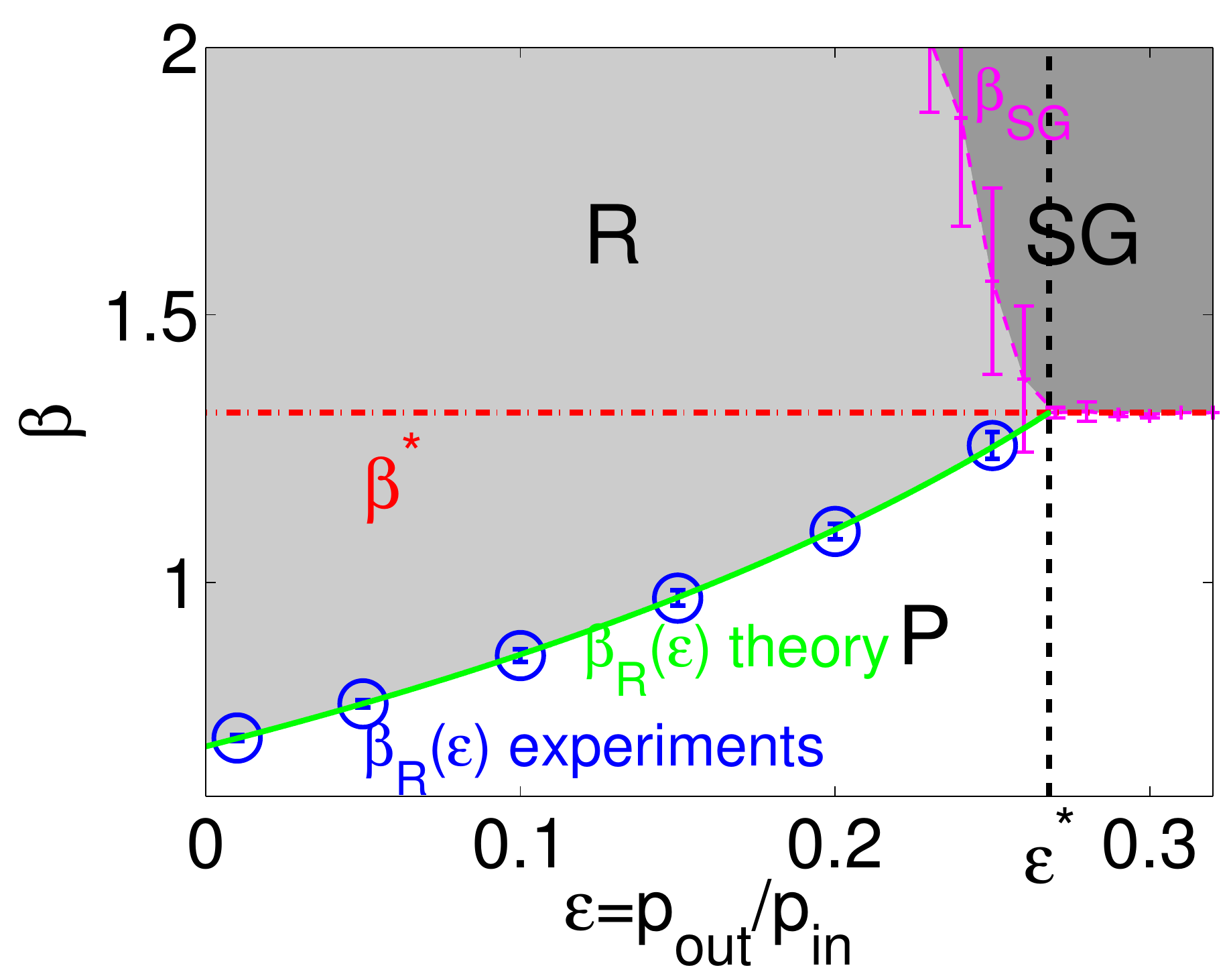} 
    \includegraphics[width=\hfw\columnwidth]{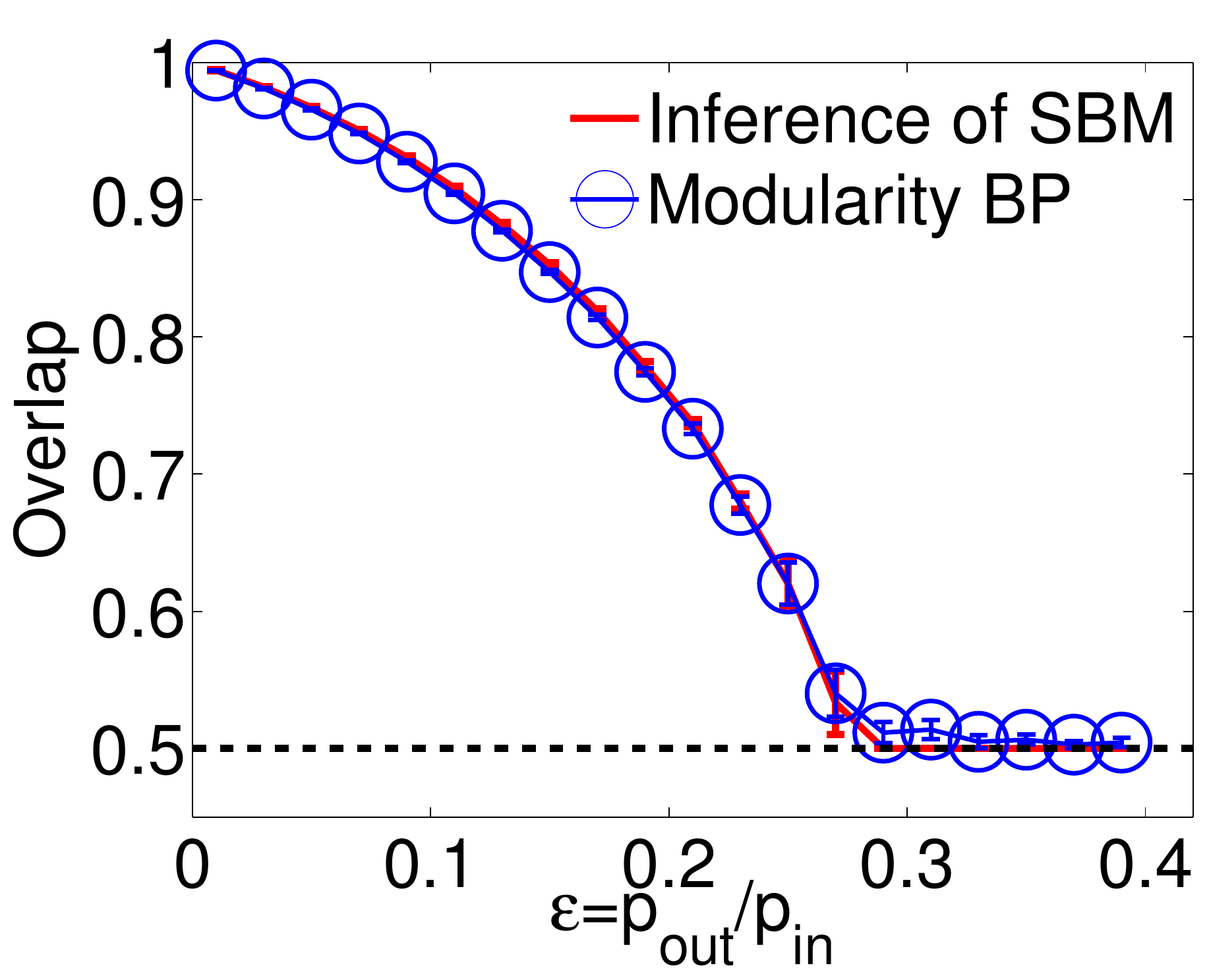} 
	\caption{
	Left: phase diagram for networks generated by the stochastic block model, showing the paramagnetic (P), retrieval (R), and spin glass (SG) phases.  Blue circles with error bars denote experimental estimates of $\betaR$, the boundary between the paramagnetic and retrieval phases, and the solid green line shows our theoretical expression~\eqref{eq:beta_eps}.  The spin glass instability occurs for $\beta > \beta^*(2,3)$ (red dash-dotted line) and $\eps^*$ is the detectability transition (black dashed line).  
	Right: The overlap of the retrieval partition at $\beta=1.315\approx\beta^*(2,3)$ (blue circles) and the partition obtained with the algorithm of~\cite{decelle-etal-pre}, which infers the parameters of the SBM with an additional EM learning algorithm.  Each experiment is on the giant component of a network with $n=10^5$, $q=2$ groups, and average degree $c=3$.  We average over $10$ random instances.\label{fig:q2}}
\end{figure}

In Fig.~\ref{fig:q2} (left) we show the phase diagram of our algorithm on SBM networks, including the paramagnetic, retrieval, and spin glass phases as a function of $\eps$, with $q=2$ and $c=3$.  The boundary $\betaR$ between the paramagnetic and retrieval phases is in excellent agreement with our expression~\eqref{eq:beta_eps}.  For $\eps < \eps^* \approx 0.267$, our algorithm finds a retrieval state for $\betaR < \beta < \betaSG$.  On the right, we show the accuracy of the retrieval partition $\{ \hat{t} \}$, defined as its \emph{overlap} with the planted partition, i.e., the fraction of nodes labeled correctly.

We emphasize that $\beta^*$ is not the optimal value of $\beta$, i.e., it is not on the Nishimori line~\cite{iba1999,NishimoriBook01,Montanari08}.  
However, the optimal $\beta$ depends on the parameters of the SBM (see Appendix).  Our claim is that setting $\beta = \beta^*$ in our algorithm succeeds throughout the easily-detectable regime, even when the parameters are unknown.  In Fig.~\ref{fig:q2} (right) we compare our algorithm with that of~\cite{decelle-etal-prl,decelle-etal-pre}, which learns the SBM parameters using an expectation-maximization (EM) algorithm.  Our algorithm provides nearly the same overlap, without the need for the EM loop.

\subsection{Results on real-world networks and choosing the number of groups} 

We tested our algorithm on a number of real-world networks.  
As for networks generated by the SBM in the detectable regime, we find a retrieval phase between the paramagnetic and spin glass phases (see figure in Appendix).  Rather than attempting to learn the optimal parameters or temperature for these networks, we simply set $\beta=\beta^*(q^*,c)$ as defined in \eqref{eq:beta-star} where $q^*$ is the ground-truth number of groups (if known) and $c$ is the average degree.  
Again, this value of $\beta$ is not optimal, and varying $\beta$ may improve the algorithm's performance; however, setting $\beta = \beta^*$ appears to work well in practice.

When the number of groups is not known, determining it is a classic model-selection problem.  The maximum modularity typically grows with $q$.  In contrast, the retrieval modularity stops growing when $q$ exceeds the correct value, giving us a principled method of choosing $q^*$ (see Appendix).  
For those networks where $q^*$ is known, we found that this procedure agrees perfectly with the ground truth. 

As shown in Table~1,
our algorithm finds a retrieval state in all these networks, with high retrieval modularity and high overlap with the ground truth.  For the Gnutella, Epinions and web-Google networks, no ground truth is known; but in contrast with~\cite{leskovec2009community}, our algorithm finds significant large-scale communities.

While most of these networks are assortative, one network in the table, the adjacency network of common adjectives and nouns in 
the novel \emph{David Copperfield}~\cite{PhysRevE.74.036104}, is disassortative, since nouns are more likely to be adjacent to adjectives than other nouns and vice versa.  In this case, we found a retrieval state with negative modularity, and high overlap with the ground truth, by setting $\beta$ to $-\beta^*(q^*,c)$. 


\label{tab:real_networks}
 \begin{table*}[!h]
\caption{Retrieval modularity, overlap between the retrieval partition and the ground truth, the number of groups $q^*$ as determined by our algorithm, the inverse temperature $\beta^*$ defined in~\eqref{eq:beta-star}, and the convergence time measured in seconds and iterations for several real-world networks~\cite{zachary1977information,lusseau2003bottlenose,PhysRevE.74.036104,polbooks,adamic2005political,leskovec2009community}.  For Gnutella, Epinions and web-Google~\cite{leskovec2009community} no ground truth is known, but based on our results we claim, contrary to~\cite{leskovec2009community}, that these networks have statistically significant large-scale communities.  
\label{tab:real_networks}
}
  \begin{tabular}{@{\vrule height 10.5pt }|l|c|c|c|c|c|c|c|c|c|c|c|}
 \hline 
 Network & $\ n\ $ & $\ m\ $ &\ $q^*\ $&\ $\beta^*$\ & $Q(\hat{t}) $ & overlap & time (sec) & \# iterations \\ 
 \hline 
 Zachary's karate club&34&78&2&1.012&0.371&1&0.001&26 \\ 
  \hline 
 Dolphin social network&62&159&2&0.948&0.395&0.887&0.001&33 \\ 
  \hline 
 Books about US politics&105&441&3&0.948 &0.521&0.829&0.002&23 \\ 
  \hline 
 Word adjacencies&112&425&2&-0.761 &-0.275&0.848&0.003&35 \\ 
  \hline 
 Political blogs&1222&16714&2&0.387 &0.426&0.948&0.043&18 \\ 
  \hline 
  Gnutella  & 62586 &147892 &7&0.995 &0.517 & &37.43&433 \\ 
 \hline
 Epinions & 75888 &405740 &4&0.632  &0.429 &&57.13&213 \\ 
 \hline
 Web-Google & 916428 &4322051&5& 0.676  &0.724 &&2331&505\\
 \hline
 \end{tabular}
 \end{table*}

\subsection{Results on hierarchical clustering}
\label{sec:hierarchy}

Many networks appear to have hierarchical structure with communities and subcommunities on many scales~\cite{clauset2004finding,clauset2008hierarchical,sales2007extracting,PhysRevE.74.036104,peixoto2013hierarchical}.  We can look for such structures by working recursively: we determine the optimal number $q^*$ of groups, divide the network into subgraphs, and apply the algorithm to each one.  We stop dividing when there is no retrieval state, indicating that the remaining subgraphs have no significant internal structure.


For networks generated by the SBM, each subgraph is an ER graph.  Our algorithm finds no retrieval state in the subgraphs, so it stops after one level of divisioin.  The same occurs in some small real-world networks, e.g.~Zachary's karate club. In some larger real-world networks, on the other hand, our algorithm repeatedly finds a retrieval state in the subgraphs, suggesting a deep hierarchical structure.  

\begin{figure}
  \centering
	\includegraphics[width=0.357\columnwidth]{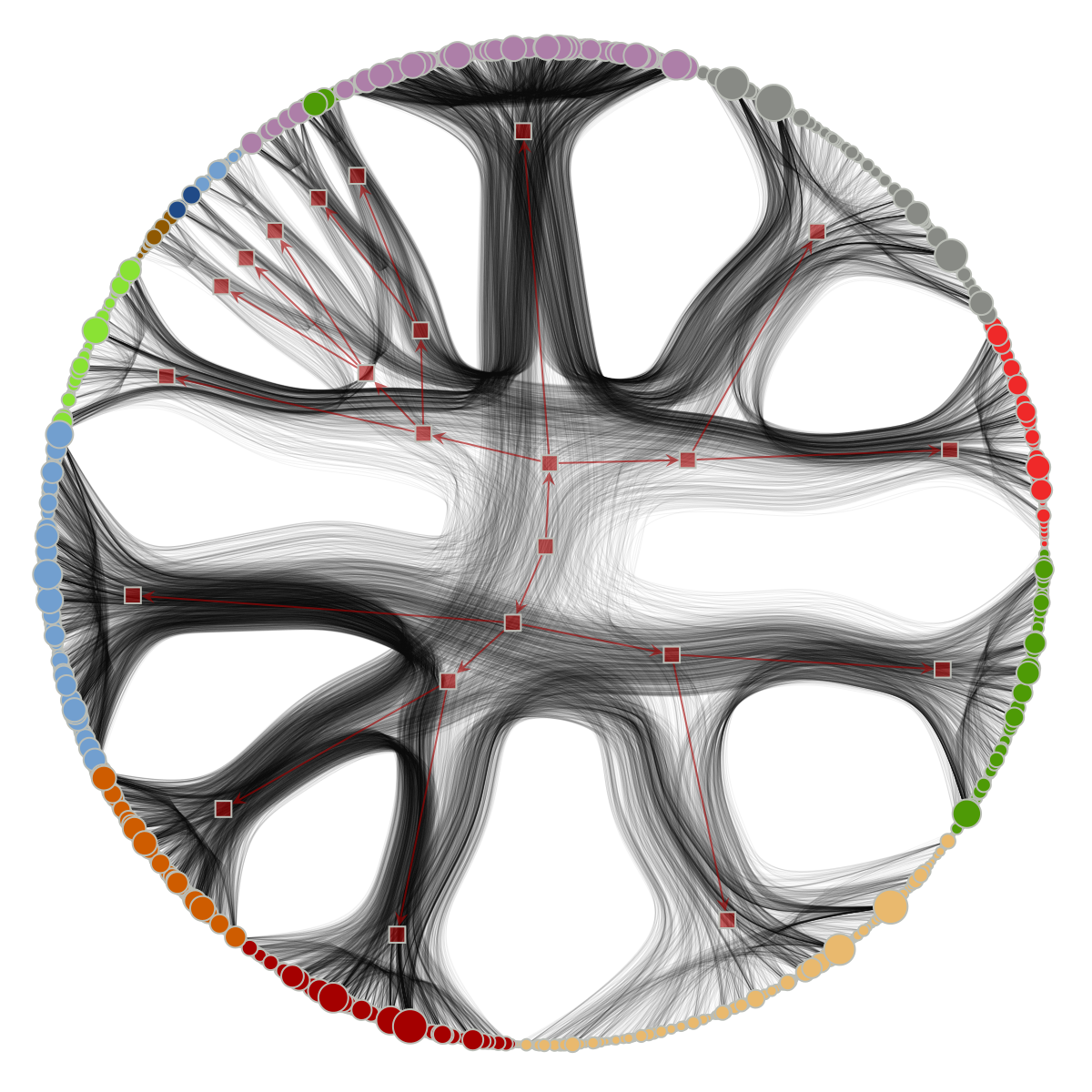}
	\includegraphics[width=0.45\columnwidth]{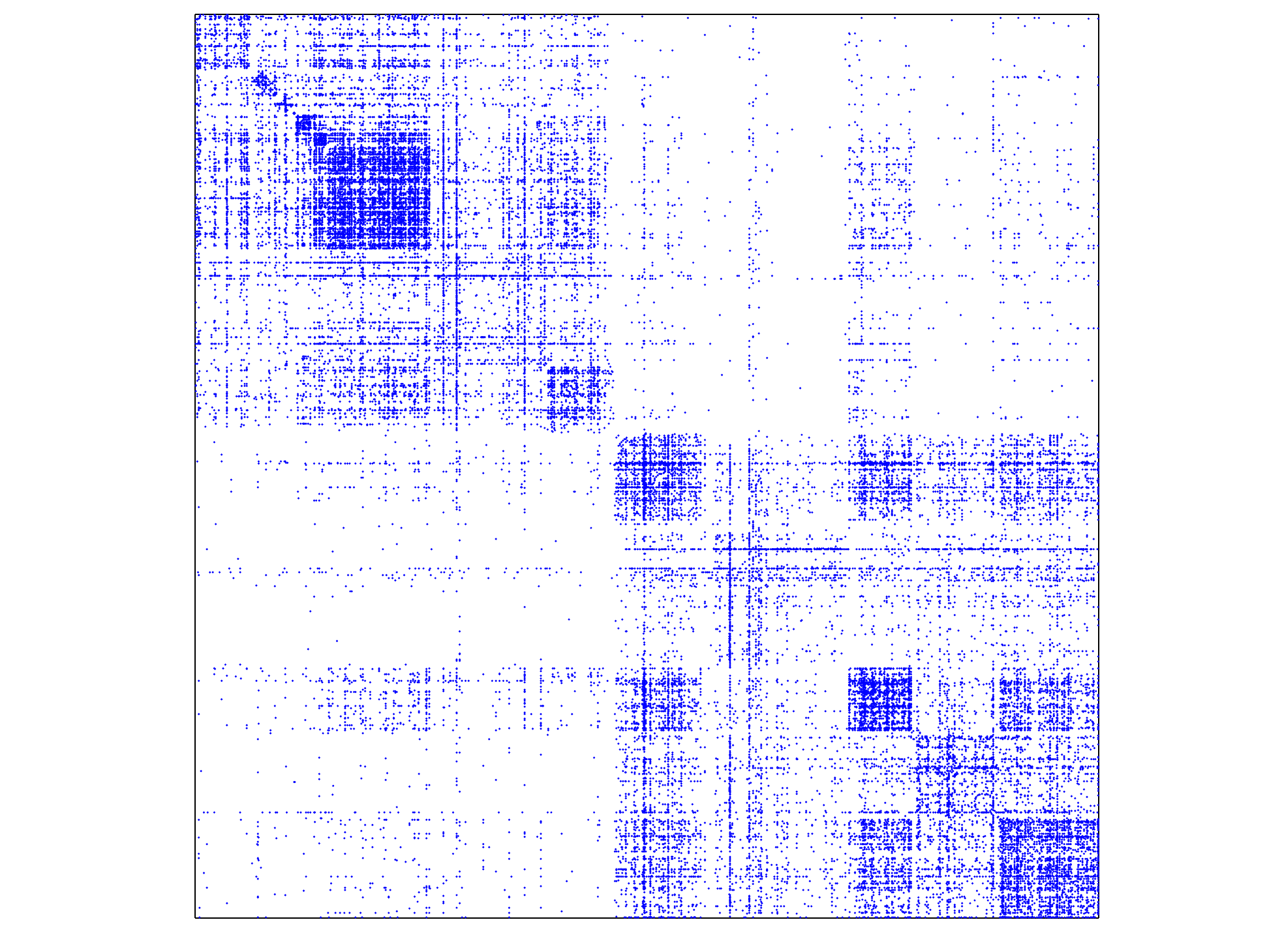}
		\caption{Left, a hierarchical division of the political blog network~\cite{adamic2005political}.   
		We apply our technique recursively, looking for a retrieval state and optimizing the 
		number of groups in which to split the community at each stage.  We stop when no retrieval 
		state is detected, indicating that the remaining groups have no statistically significant 
		subcommunities. Each leaf denotes one node, the size indicates its degree, and the colors indicate different groups in final division.   
		Right, the adjacency matrix of the network ordered according to this partition.\label{fig:polblogs_hier}}
\end{figure}

An example is the network of political blogs~\cite{adamic2005political}.  Our algorithm first finds two large communities corresponding to liberals and conservatives, and agreeing with the ground-truth labels on $95\%$ of the nodes. But as shown in Fig.~\ref{fig:polblogs_hier}, it splits these into subcommunities, eventually finding a hierarchy $5$ levels deep with a total 
of $14$ subgroups (the shaded leaves of the tree in Fig.~\ref{fig:polblogs_hier}).
We show the adjacency matrix with nodes ordered by this final partition on the right of Fig.~\ref{fig:polblogs_hier}, and the hierarchical structure is clearly visible.  The modularity of the 2nd through 5th levels are $0.426$,  $0.331$,  $0.285$, and $0.282$ respectively. This decreasing modularity may explain why the algorithm did not immediately split the network all the way down to the sub-communities.

A nested SBM was used to explore hierarchical structure in~\cite{peixoto2013hierarchical}, where the blog network was also reported to have hierarchical structure.  Our results are slightly different, giving $14$ rather than $17$ subgroups, 
but the first $3$ levels of subdivision are similar. 

\subsection{Comparison with other algorithms}
\label{sec:compare}

In this section we compare the performance of our algorithm with two popular algorithms: Louvain~\cite{blondel2008fast} and OSLOM~\cite{Lancichinetti-plosone}.  In particular, OSLOM tries to focus on statistically significant communities.  

Louvain gives partitions with similar modularity as our algorithm, but with a much larger number of groups, particularly on large networks. For example, on the Gnutella and Epinions network~\cite{leskovec2009community}, our algorithm finds $q^*=7$ and $q^*=4$ groups with modularity $0.517$ and $0.429$ respectively, while the Louvain method finds $66$ and $949$ groups with modularity $0.499$ and $0.430$ respectively.  Thus our algorithm finds large-scale communities, with a modularity similar to the smaller communities found by Louvain.  Of course, we emphasize that maximizing the modularity is not our goal: finding statistically significant communities is. 

\begin{figure}
   \centering
    \includegraphics[width=\hfw\columnwidth]{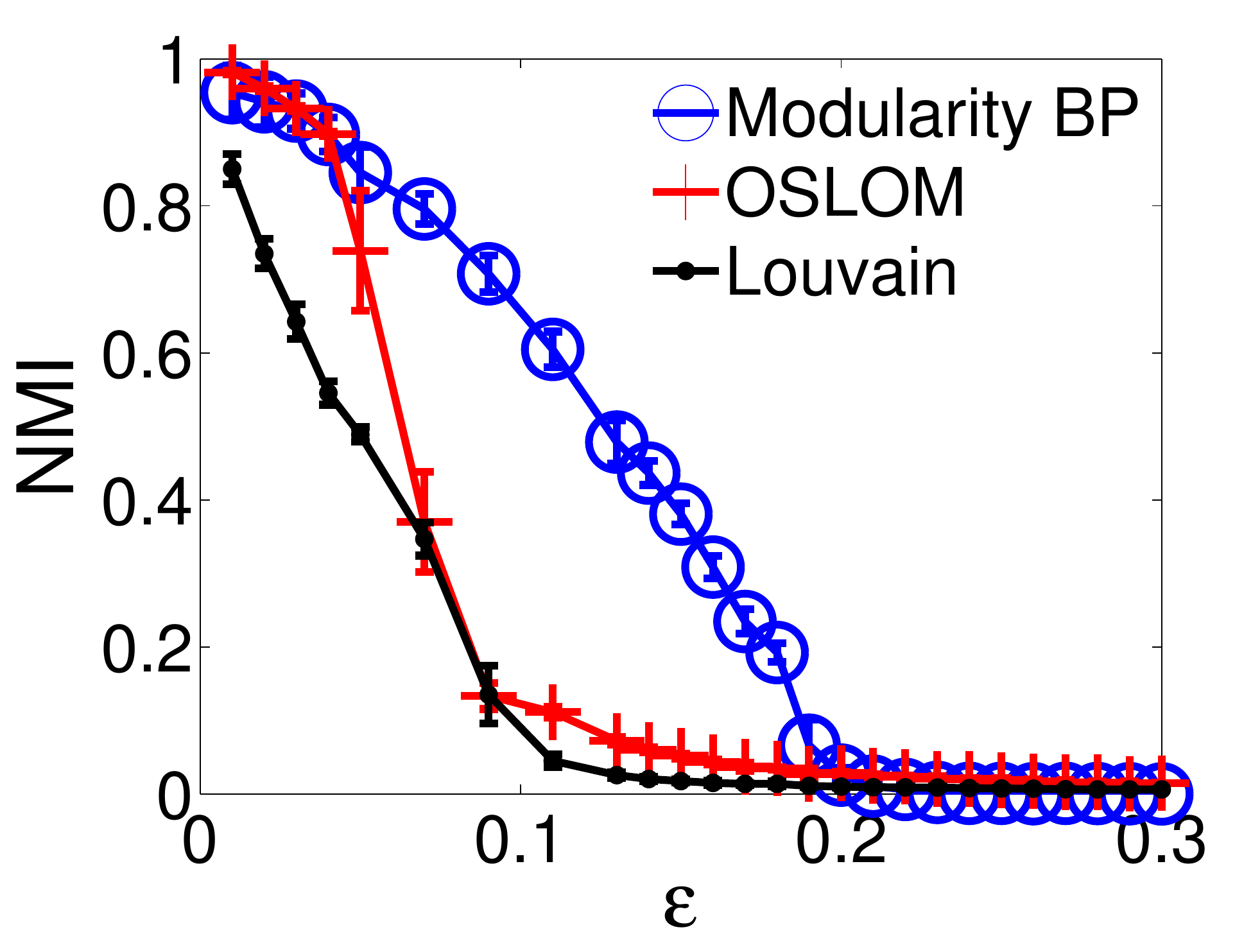} 
    \includegraphics[width=\hfw\columnwidth]{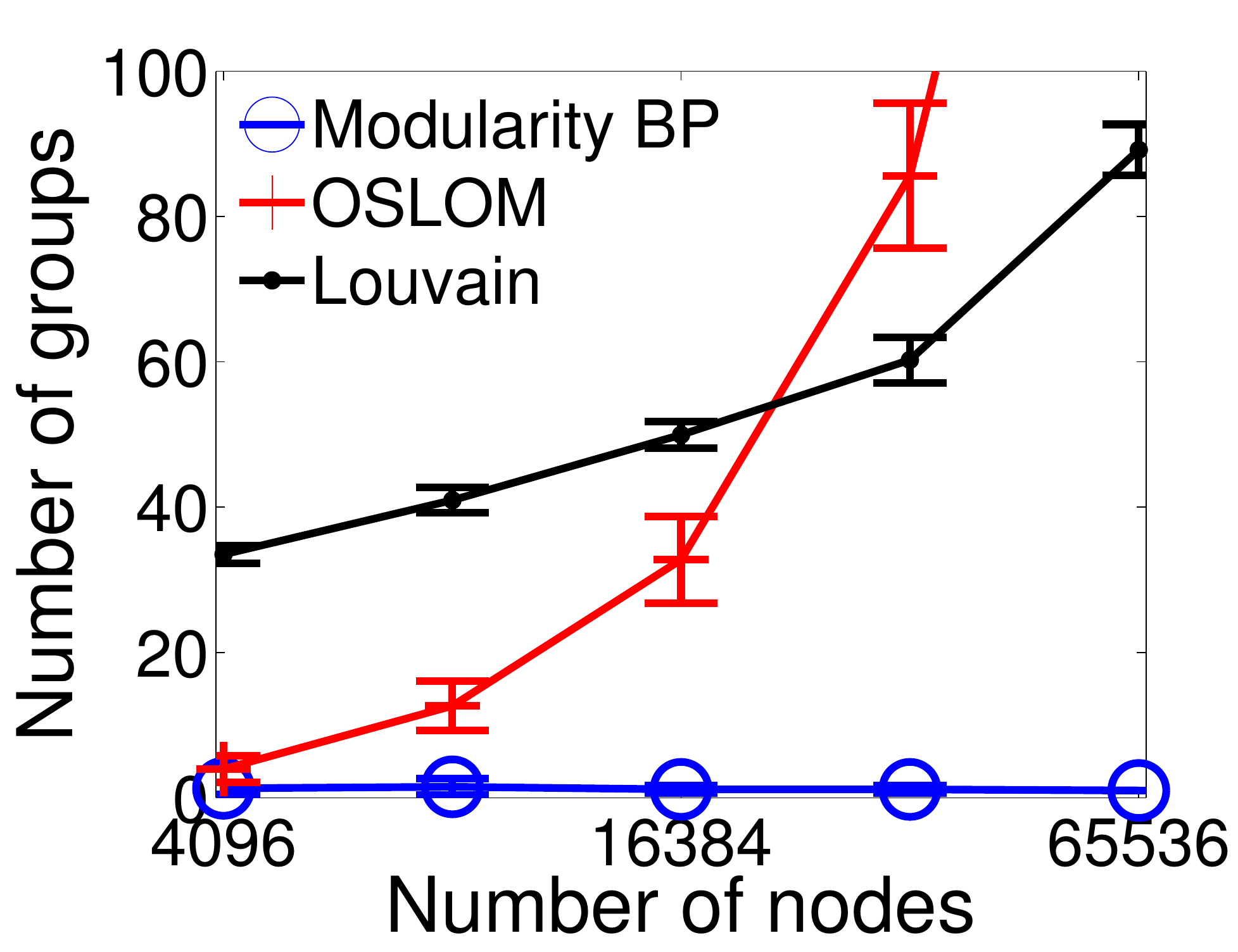} 
	\caption{Comparison of BP with Louvain and OSLOM on SBM networks with $n=10^4$, $c=6$, and $q=6$.  On the left, we show the normalized mutual information (NMI) between each algorithm's results and the true partition as a function of $\eps$; the other algorithms' NMI drops sharply well below the detectability transition at $\eps = 0.195$.  On the right, we show the inferred number of groups on the giant component of an ER graph with $c=4$.  While our algorithm correctly finds $q^*=1$, the other algorithms overfit, finding a growing number of small communities as $n$ increases.  Each point is averaged over $20$ instances.
\label{fig:compare}
	}
\end{figure}

We show results on synthetic networks in Fig.~\ref{fig:compare}.  On the left, we apply Louvain, OSLOM, and our algorithm to SBM networks with $q=6$.  We compute the normalized mutual information (NMI)~\cite{danon2005} between the inferred partition and the planted one. 
(We use the NMI rather than the overlap because the number of groups given 
by OSLOM and Louvain are very different from the planted partition.)   
For Louvain and OSLOM, the NMI drops off well below the detectability transition.  On the right, we show the number of groups that each algorithm infers for an ER graph with $c=4$.  Our algorithm correctly chooses $q=1$, recognizing that this network has no internal structure.  The other algorithms overfit, inferring a number of communities that grows with $n$.  In the Appendix we report on experiments on benchmark networks with heavy-tailed degree distributions~\cite{LFR-benchmark}, with similar results.

%

\section{Discussion}
\label{sec:discussion}

We have presented a physics-based method for finding statistically significant communities.  Rather than using an explicit generative or graphical model, it uses a popular measure of community structure, namely the modularity.  It does not attempt to maximize the modularity, which is both computationally difficult and prone to overfitting.  Instead it estimates the marginals of the Gibbs distribution using a scalable BP algorithm derived from the cavity method (see next section), and defines the retrieval partition by assigning each node to its most-likely community according to these marginals.  

In essence, the algorithm looks for the consensus of many partitions with high modularity.  When this consensus exists, it indicates statistically significant community structure, as opposed to random fluctuations.  Moreover, by testing for the existence of this retrieval state, as opposed to a spin glass state where the algorithm fluctuates between many unrelated local optima, we can determine the correct number of groups, and decompose a network hierarchically.

We note that this algorithm is related to BP for the degree-corrected stochastic block model (DCSBM).  Specifically, for a fixed $\beta$, the modularity is linearly related to the log-likelihood of the DCSBM with particular parameters (see Appendix).  However, our algorithm does not have to learn the parameters of the block model with an EM algorithm, or perform model selection between the stochastic block model and its degree-corrected variant~\cite{yan2012model}.  
To be clear, $\beta$ is still a tunable parameter that can be optimized, but the heuristic value $\beta = \beta^*$ appears to work well for a wide range of networks.

In addition to the detectability transition in the SBM, another well-known barrier to community detection is the resolution limit~\cite{Fortunato2007} where communities become difficult to find when their size is $O(\sqrt{n})$ or less.  
In the Appendix, we give some evidence that our hierarchical clustering algorithm overcomes this barrier.  Namely, for the classic example of a ring of cliques, at the second level our algorithm divides the graph precisely into these cliques.  

Another recent proposal for determining the number of groups is to use the number of real eigenvalues of the non-backtracking matrix, outside the bulk of the spectrum~\cite{Krzakala24122013}.  For some networks, such as the political blogs, this gives a larger number than the $q^*$ we found here; it may be that, in some sense, this method detects not just top-level communities, but subcommunities deeper in the hierarchy.  It would be interesting to perform a detailed comparison of the two methods.

Our approach can be extended to generalizations of the modularity, where the graph is weighted, or where a parameter $\gamma$ represents the relative importance of the expected number of internal edges~\cite{reichardt2006statistical}.  Finally, it would be interesting to apply BP to other objective functions, such as normalized cut or conductance, devising Hamiltonians from them and considering the resulting Gibbs distributions.  

Finally, we note that rather than running BP once and using the resulting marginals, we could use decimation~\cite{MMBook} to fix the labels of the most biased nodes, run BP again to update the marginals, and so on.  This would increase the running time of the algorithm, but it may improve its performance.  Another approach would be reinforcement~\cite{MMBook}, where we add external fields that point toward the likely configuration.  We leave this for future work.

\section{Methods}
\label{sec:methods}

\subsection{Defining statistical significance}
As described above, an ER random graph has many partitions with high modularity.  However, these partitions are nearly uncorrelated with each other.  In the language of disordered materials, the landscape of partitions is glassy: while the optimal one might be unique, there are many others whose modularity is almost as high, but which have a large Hamming distance from the optimum and from each other.  If we define a Gibbs distribution on the partitions, we encounter either a paramagnetic state where the marginals are uniform, or a spin glass with replica symmetry breaking where we jump between local optima.  In either case, focusing on any one of these optima is simply overfitting.

For networks such as on the right of Fig.~\ref{fig:two_A}, in contrast, there are many high-modularity partitions that are correlated with each other, and with the ground truth.  As a result, the landscape has a smooth valley surrounding the ground truth.  At a suitable temperature, the Gibbs distribution is in a retrieval phase with both low energy (high modularity) and high entropy, giving it a lower free energy than the paramagnetic state, with its marginals biased towards the ground truth.  
When BP converges to a fixed point, it finds a (local) minimum of the Bethe free energy, approximating this lower free energy phase.

We propose the existence of this retrieval phase as a physics-based definition of statistical significance.  
When it exists, the retrieval partition defined by the maximum marginals is an optimal prediction for which nodes belong to which groups.

The idea of using the free energy to separate real community structure from random noise, and using the Gibbs marginals to define a partition, also appeared in~\cite{decelle-etal-prl,decelle-etal-pre}.  However, that work is based on a specific generative model, namely the stochastic block model, and the energy is (minus) the log-likelihood of the observed network.  
In contrast, we avoid explicit generative models, and focus directly on the modularity as a measure of community structure.

\subsection{The cavity method and belief propagation}
Our goal is to compute the marginal probability distribution that each node belongs to a given group and the free energy of the Gibbs distribution.  We could do this using a Monte Carlo Markov Chain algorithm.  However, to obtain marginals we would need many independent samples, and to obtain the free energy we would need to sample at many different temperatures.  Thus MCMC is prohibitively slow for our purposes. 

Instead, for sparse networks, we can use Belief Propagation~\cite{Yedidia_etal_TR2001-22}, known in statistical physics as the cavity method~\cite{MP01}.  BP makes a conditional independence assumption, which is exact only on trees; however, in the regimes we will consider (the detectable regime of the stochastic block model, and typical real-world graphs), its estimates of the marginals are quite accurate.  It also provides an estimate of the free energy, called the Bethe free energy, which is a function of one- and two-point marginals.

BP works with ``messages'' $\psi^{i \to k}_{t}$: these are estimates, sent from node $i$ to node $k$, of the marginal probability that $t_i=t$ based on $i$'s interactions with nodes $j \ne k$.  The update equations for these messages are as follows:
\begin{equation}
\label{eq:bp}
\psi^{i \to k}_{t}
\propto 
\exp\!{\left[ - \frac{\beta d_i}{2m} \theta_t + 
\sum_{j \in \partial i \backslash k} \log \left( 1 + \psi^{j \to i}_t (\e^\beta - 1 ) \right) \right]} \, . 
\end{equation}
Here 
$\partial i$ denotes the set of $i$'s neighbors, and 
$\theta_{t}=\sum_{j=1}^n {d_j} \psi_{t}^j$ 
denotes an external field acting on nodes in group $t$, which we update after each BP iteration.  
We refer to the Appendix for detailed derivations of the BP update equations and Bethe free energy.  

For $q$ groups and $m$ edges, each iteration of~\eqref{eq:bp} takes time $O(qm)$.  If $q$ is fixed this is linear in the number of edges, and linear in the number of nodes when the network is sparse (i.e., when the average degree is constant).  Moreover, these updates can be easily parallelized.  Empirically, the number of iterations required to converge appears to depend very weakly on the network size, although in some cases it must grow at least logarithmically.  

\subsection{The factorized solution and local stability}
Observe that the \emph{factorized} solution, $\psi_{t}^{j \to i}=1/q$, where each node is equally likely to be in each possible group, is always a fixed point of~\eqref{eq:bp}.  
If BP converges to this solution, we cannot label the nodes better than chance, and the retrieval modularity is zero.  This is the paramagnetic state.

There are two other possibilities: BP fails to converge, or it converges to a non-factorized fixed point, which we call the \emph{retrieval state}.  In the latter case, we can compute the marginals by
\begin{eqnarray}
\psi^{i}_{t}
\propto 
\exp\!{\left[ - \frac{\beta d_i}{2m} \theta_t + 
\sum_{j \in \partial i } \log \left( 1 + \psi^{j \to i}_t (\e^\beta - 1 ) \right) \right]}
\, ,
\end{eqnarray}
and define the retrieval partition $\hat{t}$ that assigns each node to its most-likely community.  This partition represents the consensus of the Gibbs distribution: it indicates that there are many high-modularity partitions that are correlated with each other.  The retrieval modularity $Q(\{ \hat{t} \})$ is then a good measure of the extent to which the network has statistically significant community structure.

On the other hand, if BP does not converge, this means that neither the factorized solution nor any other fixed point is locally stable; the spin glass susceptibility diverges, and replica symmetry is broken.  In other words, the space of partitions breaks into an exponential number of clusters, and BP jumps from one to another.  The retrieval partition obtained using the current marginals will change to a very different partition if we run BP a bit longer, or if we perturb the initial BP messages slightly.  In the spin glass phase, we are free to define a retrieval modularity from the current marginals, but it fluctuates rapidly, and does not represent a consensus of many partitions.

The linear stability of the factorized solution can be characterized by computing the derivatives of messages with respect to each other at the factorized fixed point.  Using~\eqref{eq:bp}, we find that $\partial \psi^{i \to k}_{t} / \partial \psi^{j \to i}_{s} = T_{st}$ where $T_{st}$ is the $q \times q$ matrix
\begin{equation}
\label{eq:t}
T_{st} = 
	\left . \frac{\partial \psi_{t}^{i \to k}}{\partial \psi_{s}^{j \to i}} \right |_{\frac{1}{q}}
 = \frac{\e^\beta -1}{{\e^\beta}-1+q}\bra{\delta_{st} -\frac{1}{q}} \, .
\end{equation}
Its largest eigenvalue (in magnitude) is
\begin{equation}
\label{eq:lambda}
	 \lambda=\frac{\e^\beta -1}{{\e^\beta}-1+q} \, .
\end{equation}
On locally tree-like graphs with Poisson degree distributions and average degree $c$, the factorized fixed point is then unstable with respect to random noise whenever $c\lambda^2>1$.  This is also known as the de Almeida-Thouless local stability condition~\cite{AlmeidaThouless78}, the Kesten-Stigum bound~\cite{kesten_stigum_1,kesten_stigum_2}, or the threshold for census or robust reconstruction~\cite{MM06,janson2004robust}.  In our case, it shows that $\beta$ must exceed a critical $\beta^*$ given by~\eqref{eq:beta-star}.  If the network has some other degree distribution but is otherwise random, \eqref{eq:beta-star} holds where $c$ is the average excess degree, i.e., the expected number of additional neighbors of the endpoint of a random edge.

If there is no statistically significant community structure, then BP has just two phases, the paramagnetic one and the spin glass: for $\beta < \beta^*$ it converges to the factorized fixed point, and for $\beta > \beta^*$ it doesn't converge at all.  On the other hand, if there are statistically significant communities, then BP converges to a retrieval state in the range $\betaR < \beta < \betaSG$ .  Typically $\betaR < \beta^*$ and $\beta^*$ is in the retrieval phase, since even if the factorized fixed point is locally stable, BP can still converge to a retrieval state if its free energy is lower than that of paramagnetic solution.
Thus we can test for statistically significant communities by running BP at $\beta=\beta^*$.  Note that our calculation of $\beta^*$ in~\eqref{eq:beta-star} assumes that the network is random conditioned on its degree distribution; in principle $\beta^*$ could fall outside the retrieval phase for real-world networks.  
In that case, our heuristic method of setting $\beta=\beta^*$ fails, and it would be necessary to scan values of $\beta$ in the vicinity of $\beta^*$ for the retrieval state.

To estimate $\betaR$, we again consider the linear stability of BP around the factorized fixed point; but now we consider arbitrary perturbations, as opposed to random noise.  Let $T$ be the $q \times q$ matrix defined in~\eqref{eq:t}.  The matrix of derivatives of all $2qm$ messages with respect to each other is a tensor product $T \otimes B$, where $B$ is the non-backtracking matrix~\cite{Krzakala24122013}.  The adaptive external field in the BP equations suppresses eigenvectors where every node is in the same community.  As a result, the relevant eigenvalue is $\lambda \mu$ where $\lambda$ is the largest eigenvalue of $T$, and $\mu$ is the second-largest eigenvalue of $B$, and the factorized fixed point is unstable whenever $\lambda \mu > 1$.  For networks generated by the SBM, we have~\cite{Krzakala24122013} 
\begin{equation}
\label{eq:mu}
\mu = \frac{c(1-\eps)}{1+(q-1)\eps} \, . 
\end{equation}
Combining this with~\eqref{eq:lambda} and setting $\lambda \mu = 1$ gives eq.~\eqref{eq:beta_eps}. 

However, this assumes that the corresponding eigenvector of $B$ is correlated with the community structure, so that perturbing BP away from the factorized fixed point will lead to the retrieval state.  This is true as long as $\mu$ is outside the bulk of $B$'s eigenvalues, which are confined to a disk of radius $\sqrt{c}$ in the complex plane~\cite{Krzakala24122013}; if it is inside the bulk, then the community structure is washed out by isotropic eigenvectors and becomes hard to find.  Thus the communities are detectable as long as $\mu > \sqrt{c}$.  This is equivalent to $\betaR < \beta^*$, or equivalently $\eps < \eps^*$.  Thus the retrieval state exists all the way down to the Kesten-Stigum transition where $\eps = \eps^*$, $\mu = \sqrt{c}$, and $\betaR = \beta^*$.  At that point, the relevant eigenvalue crosses into the bulk, and the retrieval phase disappears.

We note that the paramagnetic, retrieval, and spin glass states were also studied in~\cite{hu2012phase}, using a generalized Potts model and a heat bath MCMC algorithm.  However, their Hamiltonian depends on a tunable cut-size parameter, rather than on a general measure of community structure such as the modularity.  Moreover, it is difficult to obtain analytical results on phase transitions using MCMC algorithms, while the stability of BP fixed points is quite tractable.

\subsection{Defining the spin glass phase}
While we have identified the spin glass phase with the non-convergence of belief propagation, the true phase diagram is potentially more complicated.  The spin glass phase is defined by the divergence of the spin glass susceptibility.  If this phase appears continuously, then in sparse problems this is equivalent to the sensitivity of the BP messages to noise, i.e., whether it converges to a stable fixed point.  However, if the spin glass phase appears discontinuously, it could be that BP converges even though the true susceptibility diverges (see e.g.~\cite{Zdeborova2009}).  

We expect this to happen above the Nishimori line when the ``hard but detectable'' phase exists~\cite{decelle-etal-pre}, when there is a retrieval state with lower free energy than the factorized fixed point but with an exponentially small basin of attraction, so that BP starting with random messages fails to converge to the true minimum of the free energy.  Detecting this spin glass phase would require us to go beyond the replica-symmetric BP equations used here to equations with one-step replica symmetry breaking~\cite{MMBook}.  In the assortative case of the stochastic block model, the hard-but-detectable phase exists for $q \ge 5$.  Happily, the corresponding range of parameters is quite narrow; nevertheless, more work on this needs to be done.

A C++ implementation can be found at \cite{code}.

\begin{acknowledgments}
We are grateful to Silvio Franz, Florent Krzakala, Mark Newman, Federico Ricci-Tersenghi, Christophe Schulke, and Lenka Zdeborov\'a for helpful discussions, and to Tiago de Paula Peixoto for drawing Fig.~\ref{fig:polblogs_hier} (left) using his software at http://graph-tool.skewed.de/.
This work was supported by AFOSR and DARPA under grant FA9550-12-1-0432.
\end{acknowledgments}


\begin{thebibliography}{39}
\expandafter\ifx\csname natexlab\endcsname\relax\def\natexlab#1{#1}\fi
\expandafter\ifx\csname bibnamefont\endcsname\relax
  \def\bibnamefont#1{#1}\fi
\expandafter\ifx\csname bibfnamefont\endcsname\relax
  \def\bibfnamefont#1{#1}\fi
\expandafter\ifx\csname url\endcsname\relax
  \def\url#1{\texttt{#1}}\fi
\expandafter\ifx\csname urlprefix\endcsname\relax\def\urlprefix{URL }\fi
\providecommand{\bibinfo}[2]{#2}
\providecommand{\eprint}[2][]{\url{#2}}

\bibitem{Luxburg07atutorial}
Von Luxburg U (2007) A tutorial on spectral clustering. Stat Comput 17:395.

\bibitem{PhysRevE.74.036104}
Newman MEJ (2006) Finding community structure in networks using the eigenvectors of matrices. Phys Rev E 74:036104.

\bibitem{Krzakala24122013}
Krzakala F, Moore C, Mossel E, Neeman J, Sly A, Zdeborov\'a L, Zhang P (2013) Spectral redemption in clustering sparse networks. Proc Natl Acad Sci USA 110:20935.

\bibitem{hastings}
Hastings MB (2006) Community detection as an inference problem. Phys Rev E 74:035102.

\bibitem{decelle-etal-pre}
Decelle A, Krzakala F, Moore C, Zdeborov\'a L (2011) Asymptotic analysis of the stochastic block model for modular networks and its algorithmic applications. Phys Rev E 84:066106.

\bibitem{decelle-etal-prl}
Decelle A, Krzakala F, Moore C, Zdeborov\'a L (2011) Inference and phase transitions in the detection of modules in sparse networks. Phys Rev Lett 107:065701.

\bibitem{karrer-newman}
Karrer B, Newman MEJ (2011) Stochastic blockmodels and community structure in networks. Phys Rev E 83:016107.

\bibitem{clauset2004finding}
Clauset A, Newman MEJ, Moore C (2004) Finding community structure in very large networks. Phys Rev E 70:066111.

\bibitem{blondel2008fast}
Blondel VD, Guillaume JL, Lambiotte R, Lefebvre E (2008) Fast unfolding of communities in large networks. J Stat Mech 2008:P10008.

\bibitem{rosvall2008maps}
Rosvall M, Bergstrom CT (2008) Maps of random walks on complex networks reveal community structure. Proc Natl Acad Sci USA 105:1118.

\bibitem{Santo201075}
Fortunato S (2010) Community detection in graphs. Physics Reports 486:75.

\bibitem{newman-girvan}
Newman MEJ, Girvan M (2004) Finding and evaluating community structure in networks. Phys Rev E 69:026113.

\bibitem{newman2004fast}
Newman MEJ (2004) Fast algorithm for detecting community structure in networks. Phys Rev E 69:066133.

\bibitem{duch2005community}
Duch J, Arenas A (2005) Community detection in complex networks using extremal optimization. Phys Rev E 72:027104.

\bibitem{guimera2004modularity}
Guimera R, Sales-Pardo M, Amaral LAN (2004) Modularity from fluctuations in random graphs and complex networks. Phys Rev E 70:025101.

\bibitem{reichardt2006statistical}
Reichardt J, Bornholdt S (2006) Statistical mechanics of community detection. Phys Rev E 74:016110.

\bibitem{zdeborova2010conjecture}
Zdeborov\'a L, Boettcher S (2010) A conjecture on the maximum cut and bisection width in random regular graphs. J Stat Mech 2010:P02020.

\bibitem{sulc-zdeborova}
Sulc P, Zdeborov\'a L (2010) Belief propagation for graph partitioning. J Phys A: Math Gen 43:B5003.

\bibitem{good2010performance}
Good BH, de~Montjoye YA, Clauset A (2010) Performance of modularity maximization in practical contexts. Phys Rev E 81:046106.

\bibitem{Lancichinetti-pre-2010}
Lancichinetti A, Radicchi F, Ramasco J (2010) Statistical significance of communities in networks. Phys Rev E 81:046110.

\bibitem{Lancichinetti-plosone}
Lancichinetti A, Radicchi F, Ramasco J, Fortunato S (2011) Finding statistically significant communities in networks. PloS One 6:e18961.

\bibitem{wilson-testing}
Wilson JD, Wang S, Mucha PJ, Bhamidi S, Nobel AB (2009) A testing based extraction algorithm for identifying significant communities in networks. Oxford University Press.

\bibitem{iba1999}
Iba Y (1999) The Nishimori line and Bayesian statistics. J Phys A: Math Gen 32:3875.

\bibitem{clauset2008hierarchical}
Clauset A, Moore C, Newman MEJ (2008) Hierarchical structure and the prediction of missing links in networks. Nature 453:98.

\bibitem{Lancichinetti-consensus}
Lancichinetti A, Fortunato S (2012) Consensus clustering in complex networks. Nature Scientific Reports 2:336.

\bibitem{mossel2012stochastic}
Mossel E, Neeman J, Sly A (2012) Stochastic block models and reconstruction. arXiv:1202.1499.

\bibitem{massoulie2013community}
Massoulie L (2013) Community detection thresholds and the weak Ramanujan property. arXiv:1311.3085.

\bibitem{mossel2013proof}
Mossel E, Neeman J, Sly A (2013) A proof of the block model threshold conjecture. arXiv:1311.4115.

\bibitem{kesten_stigum_1}
Kesten H, Stigum BP (1966) A limit theorem for multidimensional Galton-Watson processes. Ann Math Stat 37:1211.

\bibitem{kesten_stigum_2}
Kesten H, Stigum BP (1966) Additional limit theorems for indecomposable multidimensional Galton-Watson processes. Ann Math Stat 37:1463.

\bibitem{NishimoriBook01}
Nishimori H (2012) Statistical Physics of Spin Glasses and Information Processing. Oxford University Press.

\bibitem{Montanari08}
Montanari A (2008) Estimating random variables from random sparse observations. European Transactions on Telecommunications 19:385.

\bibitem{zachary1977information}
Zachary WW (1977) An information flow model for conflict and fission in small groups. Journal of Anthropological Research :452--473.

\bibitem{adamic2005political}
Adamic LA, Glance N (2005) The political blogosphere and the 2004 US election: divided they blog. Proceedings of the 3rd Intl Workshop on Link Discovery 452--473.

\bibitem{lusseau2003bottlenose}
Lusseau D, Schneider K, Boisseau OJ, Haase P, Slooten E, Dawson SM (2003) The bottlenose dolphin community of Doubtful Sound features a large proportion of long-lasting associations. Behavioral Ecology and Sociobiology 54:396.

\bibitem{polbooks}
Krebs V Social Network Analysis software \& services for organizations, communities, and their consultants, \url{www.orgnet.com/}.  Accessed September 26, 2014.

\bibitem{leskovec2009community}
Leskovec J, Lang KJ, Dasgupta A, Mahoney MW (2009) Community structure in large networks: natural cluster sizes and the absence of large well-defined clusters. Internet Math 6:29.

\bibitem{sales2007extracting}
Sales-Pardo M, Guimera R, Moreira AA, Amaral LAN (2007) Extracting the hierarchical organization of complex systems. Proc Natl Acad Sci USA 104:15224.

\bibitem{peixoto2013hierarchical}
Peixoto TP (2014) Hierarchical block structures and high-resolution model selection in large networks. Phys Rev X 4:011047.

\bibitem{danon2005}
Danon L, Diaz-Guilera A, Duch J, Arenas A (2005) Comparing community structure identification. J Stat Mech 2005:P09008.

\bibitem{LFR-benchmark}
Lancichinetti A, Fortunato S, Radicchi F (2008) Benchmark graphs for testing community detection algorithms. Phys Rev E 78:046110.

\bibitem{yan2012model}
Yan X, Jensen JE, Krzakala F, Moore C, Shalizi CR, Zdeborov\'a L, Zhang P, Zhu Y (2014) Model selection for degree-corrected block models. J Stat Mech 2014:P05007.

\bibitem{Fortunato2007}
Fortunato S, Barthelemy M (2007) Resolution limit in community detection. Proc Natl Acad Sci USA 104:36.

\bibitem{Yedidia_etal_TR2001-22}
Yedidia J, Freeman W, Weiss Y (2003) Understanding belief propagation and its 
generalizations. Exploring Artificial Intelligence in the New Millennium 
(Morgan Kaufmann Publishers Inc., San Francisco).

\bibitem{MP01}
M\'ezard M, Parisi G (2001) The Bethe lattice spin glass revisited. Eur Phys J B 20:217.

\bibitem{AlmeidaThouless78}
De~Almeida J, Thouless D (1978) Stability of the Sherrington-Kirkpatrick solution of a spin glass model. J Phys A: Math Gen 11:983.

\bibitem{MM06}
M\'ezard M, Montanari A (2006) Reconstruction on trees and spin glass transition. J Stat Phys 124:1317.

\bibitem{janson2004robust}
Janson S, Mossel E (2004) Robust reconstruction on trees is determined by the second eigenvalue. Ann Prob :2630--2649.

\bibitem{hu2012phase}
Hu D, Ronhovde P, Nussinov Z (2012) Phase transitions in random Potts systems and the community detection problem. Phil Mag 92:406.

\bibitem{Zdeborova2009}
Zdeborov\'a L (2009) Statistical physics of hard optimization problems. Acta Phys Slov 59:169.

\bibitem{MMBook}
M\'ezard M, Montanari A (2009) Information, Physics, and Computation. Oxford University Press.

\bibitem{code} 
A C++ implementation of our algorithm can be found at \url{http://panzhang.net}.


\end{thebibliography}

\begin{appendix}
\section{Belief Propagation equation and Bethe free energy}
\label{eq:si-bp}

In this section we derive the BP update equations appearing in the main text.  BP works with ``messages'' $\psi^{i\to k}_{t}$: these are estimates, sent from node $i$ to node $k$, of the marginal probability that $t_i=t$ based on $i$'s interactions with nodes $j \ne k$.  If the Hamiltonian is $-mQ$, the update equations for these messages are as follows:
\begin{align}
\label{eq:bp0}
	\psi^{i\to k}_{t}
	&= \frac{1}{Z_{i\to k}} \prod_{j\in\partial i\backslash k}
	\sum_{s=1}^q \e^{\beta\delta_{st}} \psi^{j\to i}_{s}
	\prod_{j\neq i,k} \sum_{s=1}^q \e^{-\beta\frac{d_id_j}{2m}\delta_{st}}\psi^{j\to i}_{s}  \nonumber \\
	&= \frac{1}{Z_{i\to k}} \prod_{j \in \partial i \backslash k} \left( 1 + \psi^{j \to i}_t (\e^\beta - 1 ) \right)
	\prod_{j\neq i,k} \left( 1 + \psi^{j \to i}_t (\e^{-\beta\frac{d_id_j}{2m}} - 1) \right) \, .
\end{align}
Here $Z_{i\to k}$ is simply a normalization factor, and $\partial i$ denotes the neighborhood of node $i$.  The BP estimate of the marginal probability $\psi^i_t = \Pr[t_i=t]$ is then
\begin{align}
	\psi^{i}_{t}
	&= \frac{1}{Z_{i}} \prod_{j\in\partial i}
	\sum_{s=1}^q \e^{\beta\delta_{st}}\psi^{j\to i}_{s} 
	\prod_{j\neq i} \sum_{s=1}^q \e^{-\beta\frac{d_id_j}{2m}\delta_{st}}\psi^{j\to i}_{s} \nonumber \\
	&= \frac{1}{Z_{i}} \prod_{j\in\partial i}
	\left( 1 + \psi^{j \to i}_t (\e^\beta - 1 ) \right)
	\prod_{j\neq i} \left( 1 + \psi^{j \to i}_t (\e^{-\beta\frac{d_id_j}{2m}} - 1) \right) \, , \label{eq:marginals}
\end{align}
which is the same as~\eqref{eq:bp0} except that we remove the condition $j \ne k$.  We can also estimate the two-point marginals, and in particular, the probability that two neighboring points belong to the same group.  If $\brc{ij} \in \edges$, the BP estimate of the probability that $t_i=t$ and $t_j=s$ is 
\begin{equation}
\label{eq:two-point}
\psi^{ij}_{st} = \frac{1}{Z_{ij}} \e^{\beta \delta_{st}} \psi^{j \to i}_{s} \psi^{i \to j}_{t} \, . 
\end{equation}

The update equations~\eqref{eq:bp0} involve $qn^2$ messages: every node interacts with every other one, not just their neighbors.  However, in the sparse case we can simplify the effect of non-neighbors, by replacing them with an external field as in~\cite{decelle-etal-prl,decelle-etal-pre}.  If 
$k \notin \partial i$ and $d_i, d_k \ll \sqrt{m}$, we have
\[
\psi^{i}_{t} 
= \psi^{i\to k}_{t} \sum_{s} \e^{-\beta\frac{d_id_k}{2m}\delta_{st}}\psi^{k\to i}_{s}
\approx \psi^{i\to k}_{t} \bra{1-\beta\frac{d_id_k}{2m}\psi^{k\to i}_{t}} 
\approx \psi^{i\to k}_{t} \, . 
\]
In that case, we can identify the messages $\psi^{i \to k}_t$ that $i$ sends to its non-neighbors $k$ with its marginal $\psi^i_t$.  Then~\eqref{eq:bp0} simplifies to
\begin{align}
\psi^{i\to k}_{t} 
&= \frac{1}{Z_{i\to k}} \prod_{j \in \partial i \backslash k} \left( 1 + \psi^{j \to i}_t (\e^\beta - 1 ) \right)
	\prod_{j\neq i,k} \left( 1 + \psi^{j}_t (\e^{-\beta\frac{d_id_j}{2m}} - 1) \right) \nonumber \\ 
&\approx 	\frac{1}{Z_{i\to k}} 
\,\exp\!\left( - \frac{\beta d_i}{2m} \theta_t + 
\sum_{j \in \partial i \backslash k} \log \left( 1 + \psi^{j \to i}_t (\e^\beta - 1 ) \right) \right)
	\, , \label{eq:bp}
\end{align}
where
\begin{equation}
	\theta_{t}=\sum_{j=1}^n {d_j} \psi_{t}^j
\end{equation} 
denotes an external field acting on nodes in group $t$, which we update after each BP iteration.  Iterating~\eqref{eq:bp} now has computational complexity $qm$, which is linear in the number of edges when $q$ is fixed.  


The \emph{Bethe free energy} of a BP fixed point is a function of the messages:
\begin{equation}
	\fBethe = -\frac{1}{n\beta}\bra{\sum_i \log Z_i - \sum_{\brc{ij}\in\edges}\log Z_{ij} +\frac{\beta}{4m}\sum_t \theta_{t}^2 } \, , 
\end{equation}
where $Z_i$ and $Z_{ij}$ are the normalization constants for the one- and two-point marginals appearing in~\eqref{eq:marginals} and~\eqref{eq:two-point}.   BP fixed points are also stationary points of the Bethe free energy~\cite{Yedidia_etal_TR2001-22}.

\begin{figure}
   \centering
\includegraphics[width=\fw\columnwidth]{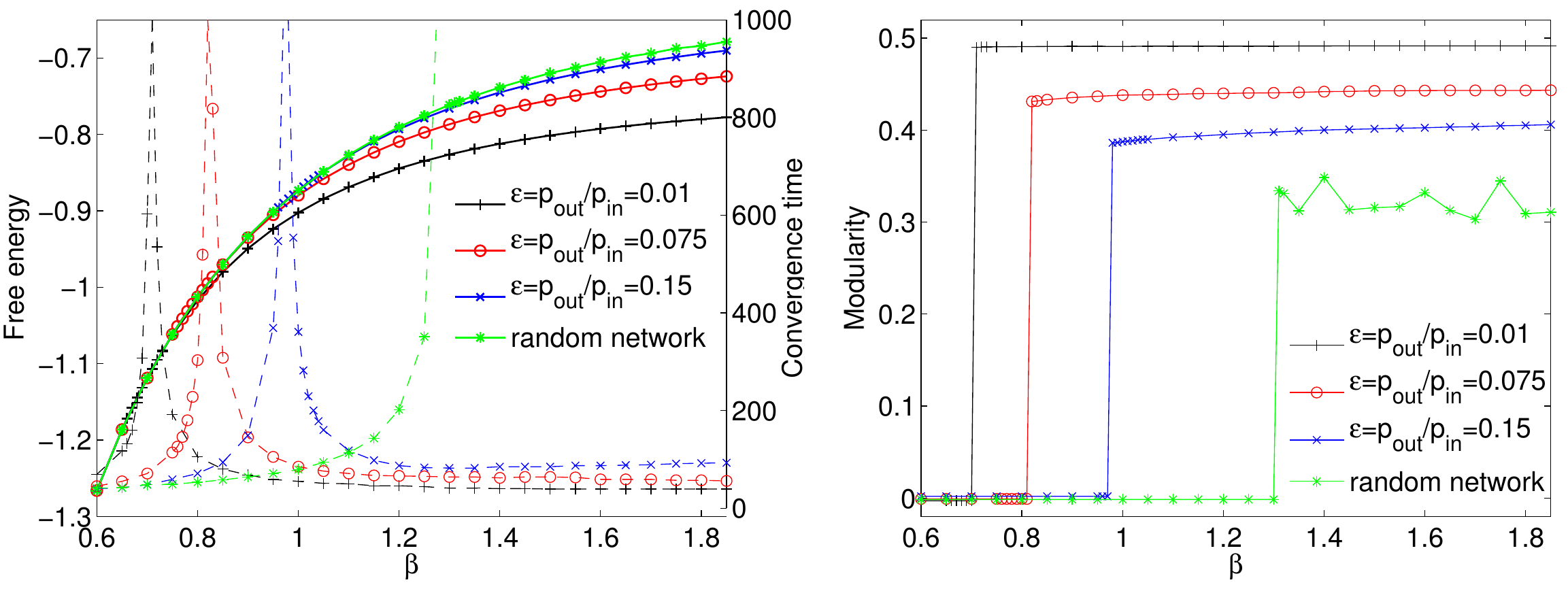}
\caption{Left: Free energy (solid) and convergence time (dashed) as a function of $\beta$ for networks generated by the stochastic block model for three different values of $\eps=\cout/\cin$, also compared with an Erd\H{o}s-R\'enyi graph.  Right: retrieval modularity for these networks.  All networks have size $n=10^4$ and average degree $c=3$.  The networks generated by the SBM have $q=2$ groups of equal size.\label{fig:f_time}}
\end{figure}

Observe that the \emph{factorized} solution, $\psi_{t}^{j\to i}=1/q$, where each node is equally likely to be in each possible group, is always a fixed point of the BP equations~\eqref{eq:bp}.  
Assuming it does not get stuck in a local minimum, BP converges to a retrieval state whenever its Bethe free energy is less than that of the factorized state.  If the network has average degree $c$, this is simply
\[
	\fBethe^{\textrm{fact}}
	= -\frac{1}{\beta} \bra{\log q+\frac{c}{2} \log\left( 1-\frac{1}{q}+\frac{\e^\beta}{q} \right) - \frac{c\beta}{2q}} \, . 
\]
In Fig.~\ref{fig:f_time} we compare the free energy, convergence time, and retrieval modularity for networks generated by the stochastic block model at three different values of $\eps$, alongside an Erd\H{o}s-R\'enyi graph of the same average degree $c=3$.  For small enough $\beta$, their free energies are all equal to $\fBethe^{\textrm{fact}}$, since they are all in the paramagnetic phase.  For each value of $\eps$, there is a critical $\betaR$ at which the free energy splits off from the others, where makes a transition to a retrieval state with $\fBethe < \fBethe^{\textrm{fact}}$.  The retrieval modularity jumps to a nonzero value, indicating community structure, and the convergence time diverges at the transition.  For the Erd\H{o}s-R\'enyi graph, the apparent modularity also jumps, but at $\beta^* = \betaSG$ it enters the spin glass phase rather than the retrieval phase: BP fails to converge and the retrieval modularity fluctuates, indicating partitions that are uncorrelated with each other.

\section{Relation with the degree-corrected stochastic block model}
\label{sec:si-dcsbm}



The degree-corrected stochastic block model (DCSBM) was introduced in~\cite{karrer-newman} to overcome the fact that the SBM typically places low-degree and high-degree vertices into different groups, since it expects the degree distribution within each group to be Poisson.  The DCSBM's parameters are the expected node degrees $\{d_i\}$ and a $q \times q$ matrix of parameters $\omega_{rs}$.  Given a partition $\{t\}$, the number of edges $A_{ij}$ between each pair $\brc{ij}$ is Poisson-distributed with mean $d_i d_j \omega_{t_i,t_j}$.  In the simple graph case where $A_{ij}=1$ if $\brc{ij} \in \edges$ and $A_{ij}=0$ otherwise, the log-likelihood of the network is then
\begin{align}
	L(\{t\})
	&= \log P(G|\{\omega_{ab}\},\{t\}) \nonumber \\
	&= \log\bra{\prod_{\brc{ij} \in \edges} d_i d_j \omega_{t_it_j} \prod_{\brc{ij}} \e^{-d_i d_j \omega_{t_it_j}}} \, . 
\label{eq:dcsbm-like}	
\end{align}
If $\omega_{rs}=\omegain$ for $r=s$ and $\omegaout$ for $r \ne s$, the likelihood can be written as
\begin{align}
	L &= \sum_{\brc{ij}} \big( \log (d_i d_j \omegaout) - d_id_j\omegaout \big) 
	+ \left( \log \frac{\omegain}{\omegaout} \right ) 
	\brb {\sum_{\brc{ij}\in\edges}\delta_{t_it_j}
	- \frac{\omegain-\omegaout}{\log (\omegain/\omegaout)} \sum_{\brc{ij}} d_i d_j \delta_{t_it_j}}.
\label{eq:dcsbm}
\end{align}
Comparing with the definition of modularity, if we set $\omegain$ and $\omegaout$ such that  
\begin{equation}
\beta=\log \frac{\omegain}{\omegaout}
\quad \text{and} \quad
2m = \frac{\log (\omegain/\omegaout)}{\omegain-\omegaout} \, ,
\label{eq:2m}
\end{equation}
then the second term in~\eqref{eq:dcsbm} is $\beta m Q(\{t\})$.  Since the first term in~\eqref{eq:dcsbm} does not depend on $\{t\}$, we have
\[
\e^{L(\{t\})} \propto \e^{\beta m Q(\{t\})} \, ,
\]
and the Gibbs distribution is exactly the Gibbs distribution of partitions in the DCSBM.  

Thus, for any fixed $\beta$, there are parameters $\omegain, \omegaout$ of the DCSBM such that these distributions have the same free energy and the same ground state.  Belief propagation on the DCSBM was described in~\cite{yan2012model}, and one can optimize the parameters $\omegain, \omegaout$ through an expectation-maximization algorithm analogous to~\cite{decelle-etal-pre,decelle-etal-prl}.  However, our approach is different in several ways.
\begin{itemize}
\item{We define community structure directly in terms of a classic measure, the modularity, as opposed to the log-likelihood of a generative model.}
\item{Rather than having to fit the parameters of the DCSBM with an EM algorithm, we have a single temperature parameter $\beta$.  We can usually detect communities by setting $\beta = \beta^*$ as in main text; at worst, we just have to a scan a small region.}  
\item{For real-world networks the retrieval modularity appears to be a good guide to the number of groups $q^*$, while the free energy of the (DC)SBM continues to decrease for $q > q^*$.}
\item{Our approach appears to work equally well for networks with Poisson degree distributions (generated by the SBM) and those with heavy-tailed degree distributions, such as the LFR benchmark~\cite{LFR-benchmark} and the network of political blogs, where the DCSBM does much better~\cite{karrer-newman}.  In particular, we have no need to do model selection between SBM and DCSBM, as was done using the Bethe free energy in~\cite{yan2012model}.}
\end{itemize}

\section{The Nishimori line and the optimal temperature} 
\label{sec:si-nishimori}

When data is produced by an underlying generative model, inference of the latent parameters can be done optimally along the 
Nishimori line~\cite{iba1999,NishimoriBook01}, where the Gibbs distribution is exactly the posterior distribution of the latent parameters (in this case the group labels or partitions).  If the network is generated by the DCSBM, then~\eqref{eq:2m} gives a $\betanish$ that corresponds to the correct parameters at Nishimori line.  Determining the parameters, and therefore $\betanish$, could be done with an EM algorithm as in~\cite{decelle-etal-prl,decelle-etal-pre}, but our goal is to avoid this additional learning step.  Moreover, if the network is not actually generated by the DCSBM, there is a priori no value of $\beta$ that corresponds to the Nishimori line, and no way to determine the optimal $\beta$ without access to the ground truth.

\begin{figure}
   \centering
    \includegraphics[width=\fw\columnwidth]{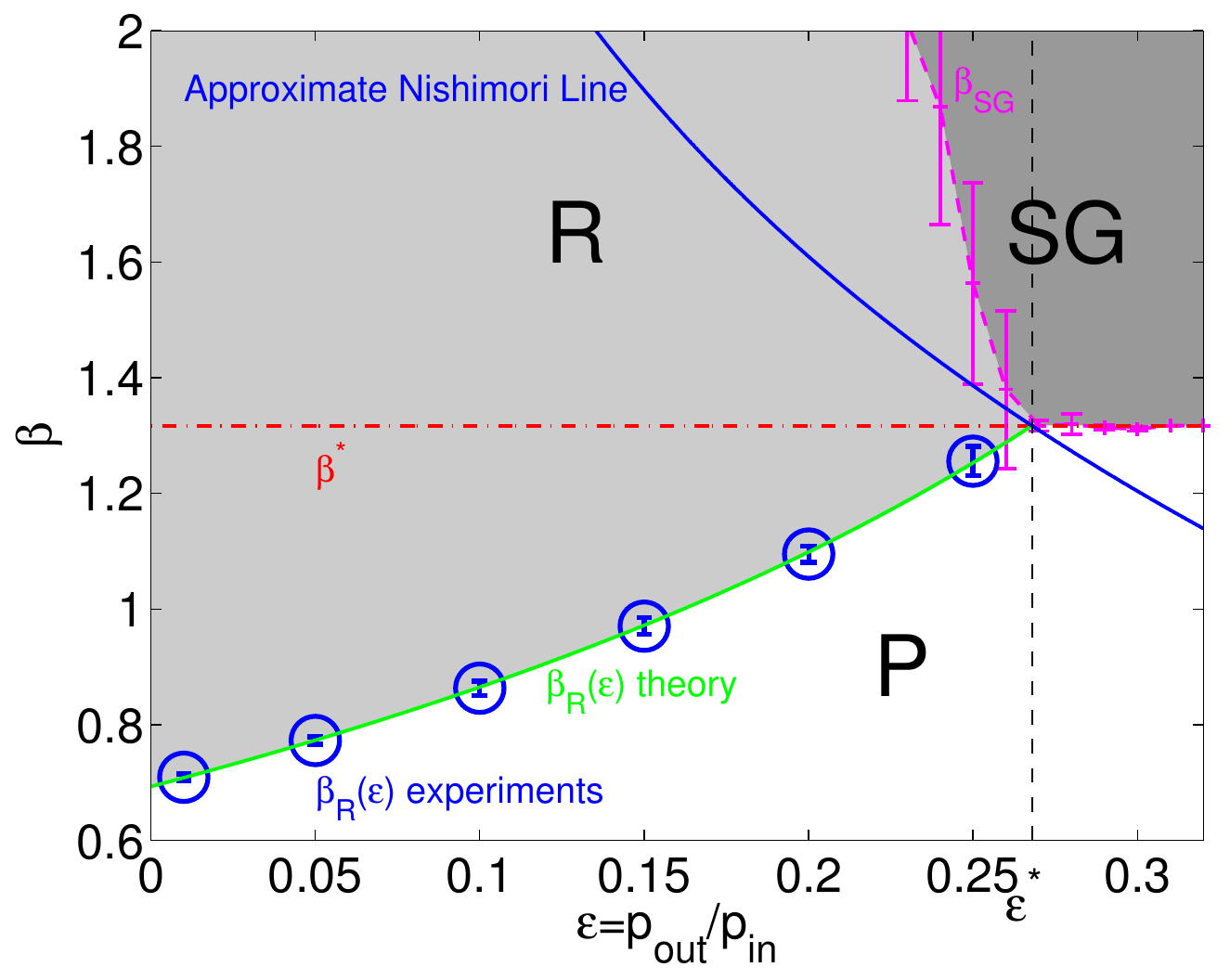} 
\caption{
The phase diagram from the main text for networks generated by the stochastic block model, with the approximate Nishimori line $\betanish = -\log \eps$ added (blue).  Replica symmetry breaking cannot occur on the Nishimori line, and indeed it avoids the spin-glass phase.  Inference at $\betanish$ would be optimal, but it would require us to learn, or infer, the correct value of the parameter $\eps$.  
\label{fig:nishimori}}
\end{figure}

However, for synthetic networks generated by the SBM, we can construct an approximate Nishimori line by omitting the difference between the SBM and the DCSBM, by assuming that the expected degrees are actually the same.  This gives
\[
\betanish = \log(\cin/\cout) = -\log \eps \, . 
\]
In Fig.~\ref{fig:nishimori} we show the phase diagram from the main text with this approximate Nishimori line added.  It passes through the critical point $(\eps^*, \beta^*)$ (one can check analytically that $\beta^* = -\log \eps^*$) and that it avoids the spin-glass phase, passing directly from the paramagnetic phase to the retrieval phase.  This recovers the fact that replica symmetry breaking cannot occur on the Nishimori line~\cite{Montanari08}.

\section{Choosing the number of groups}
\label{sec:choosing}

Choosing the number $q$ of groups in a network is a classic model selection problem.  Setting $q$ by maximizing the modularity is a widely-used heuristic in the network literature; however, as we have already seen, it is prone to overfitting.  For example, the maximum modularity for an Erd\H{o}s-R\'enyi graph is an increasing function of $q$, while the correct model has $q=1$.  Similarly, in the stochastic block model the likelihood increases, or the ground state energy decreases, until every node is assigned to its own group.

One approach~\cite{decelle-etal-prl,decelle-etal-pre} is to use the free energy rather than the ground state energy.  In essence, the entropic term penalizes overfitting, and gives us the total likelihood of the model summed over all partitions, as opposed to the likelihood of the best partition.  This approach works well on synthetic graphs: the free energy decreases until we reach the correct number of groups, after which it stays roughly constant.  However, on real-world networks the free energy continues to decrease with $q$, for example as shown in Fig.~8 of~\cite{decelle-etal-pre}.  Thus, for networks not generated by the SBM, it is not clear that this method works.

\begin{figure}
   \centering
\includegraphics[width=\fw\columnwidth]{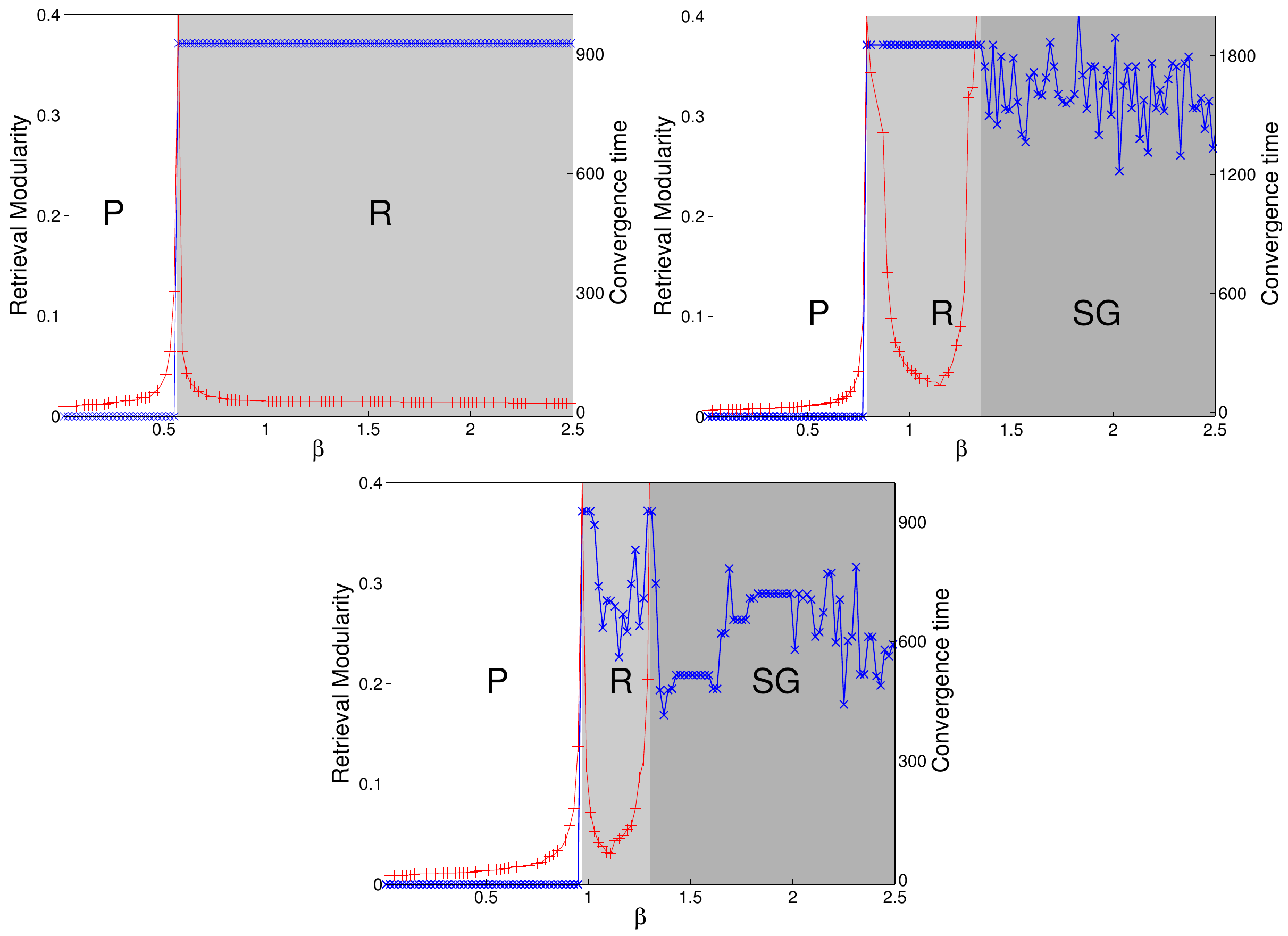} 
\caption{Retrieval modularity (blue $\times$) and BP convergence time (red $+$) of
Karate club network with $2$ groups (top left), $3$ groups (top right), and $4$ groups (bottom). 
With $q=2$, which is the ground truth value, the system has a very strong community structure, represented by a large retrieval phase  
starting at $\betaR = 0.565$.  
With $q=3$, the retrieval phase exists between $\betaR = 0.79$ and $\betaSG = 1.35$; compare Fig.~2 (right) in the main text.  
With $q=4$ groups, the retrieval phase becomes even narrower, between $\betaR = 0.97$ and $\betaSG = 1.3$.
\label{fig:phases_karate}}
\end{figure}

Here we propose to use the retrieval modularity $Q(\{\hat{t}\})$ as a criterion for choosing $q$.  Namely, we claim that $Q(\{\hat{t}\})$ increases with $q$ until we reach the correct value $q^*$.  For $q > q^*$, either $Q(\{\hat{t}\})$ stays the same, or the retrieval phase disappears and we enter the spin glass phase.  In Fig.~\ref{fig:phases_karate} we plot $Q(\{\hat{t}\})$ and BP convergence time for the karate club network with different values of $q$. With $q=2$, i.e., the ground-truth number of groups, the retrieval phase is very large.  For larger $q$, the retrieval phase becomes narrower, and $Q(\{\hat{t}\})$ does not increase.  Note the similarity with Fig.~2 (right) in the main text.  

\begin{figure}
   \centering
\includegraphics[width=\fw\columnwidth]{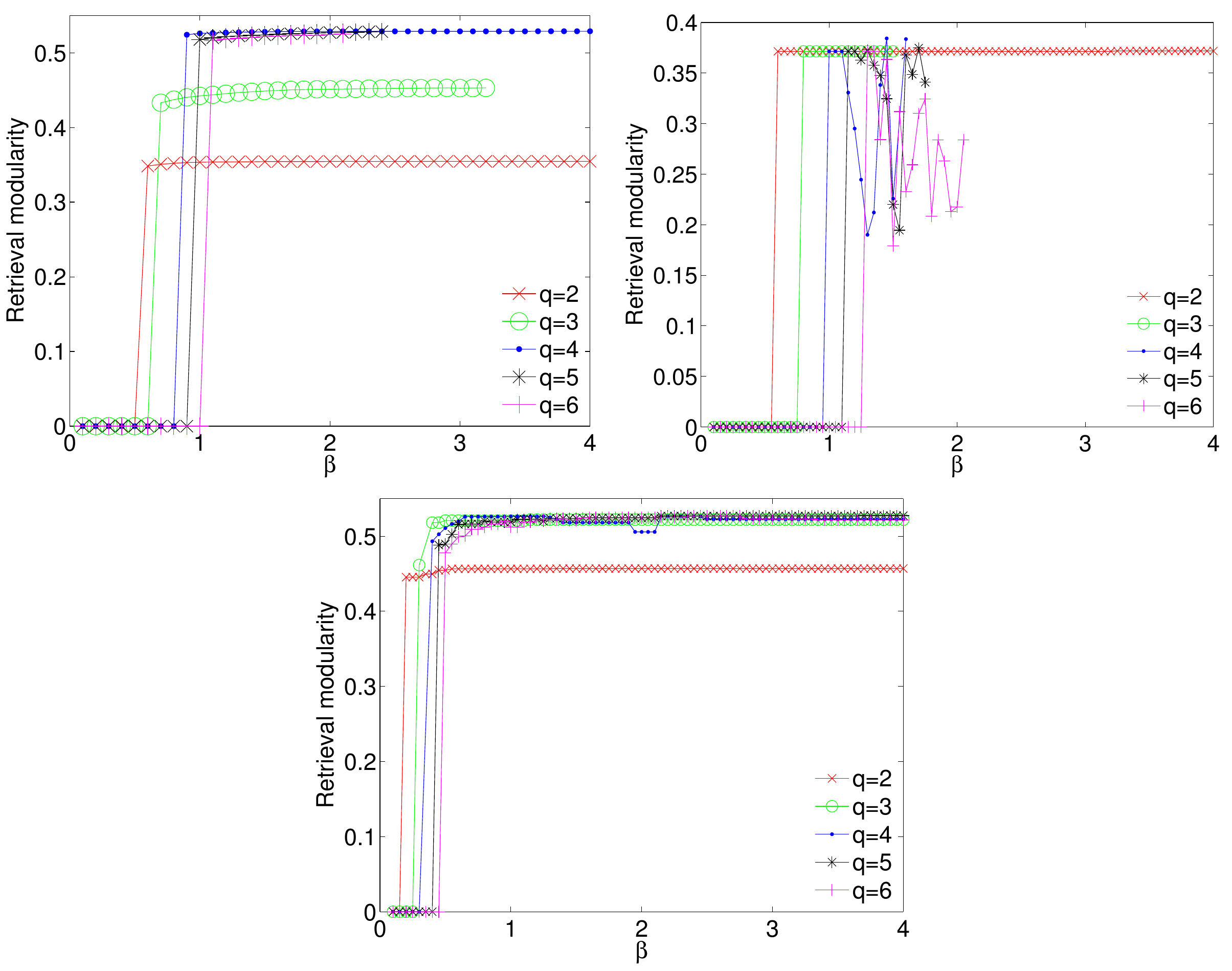} 
	\caption{Retrieval modularity as a function of $q$ for three networks where the number of groups is known: a network generated by the 
	stochastic block model with $q^*=4$, $n=10^4$, and $\eps=0.1$ (top left), the karate club with $q^*=2$ (top right) and the network of political 
	books with $q^*=3$ (bottom). In each case, for $q > q^*$ the retrieval modularity stops growing until the spin glass phase appears.  \label{fig:real_q}}
\end{figure}

In Fig.~\ref{fig:real_q}, we plot $Q(\{\hat{t}\})$ for different values of $q$ as a function of $\beta$ for three networks with known community structure: a synthetic network generated by the SBM with $q^*=4$, the karate club with $q^*=2$~\cite{zachary1977information}, and a network of political books with $q^*=3$~\cite{polbooks}.  In each case, $Q(\{\hat{t}\})$ stops growing at $q=q^*$, and is nearly independent of $\beta$ throughout the retrieval phase.  
(To deal with fluctuations, in practice we do not increase q unless the retrieval modularity increases by at least some threshold value.)
Thus our method gives the correct number of communities, rather than overfitting. 

Note that here $q^*$ refers to the top level of organization in the network.  In the main text, we discuss using our approach to recursively divide communities into subcommunities.  In that case, we use this procedure to determine the number $q^*$ of subcommunities we should split the network into at each stage, and stop splitting when we reach communities with $q^*=1$.  

\section{Additional comparisons with Louvain and OSLOM}

\begin{figure}
   \centering
\includegraphics[width=\fw\columnwidth]{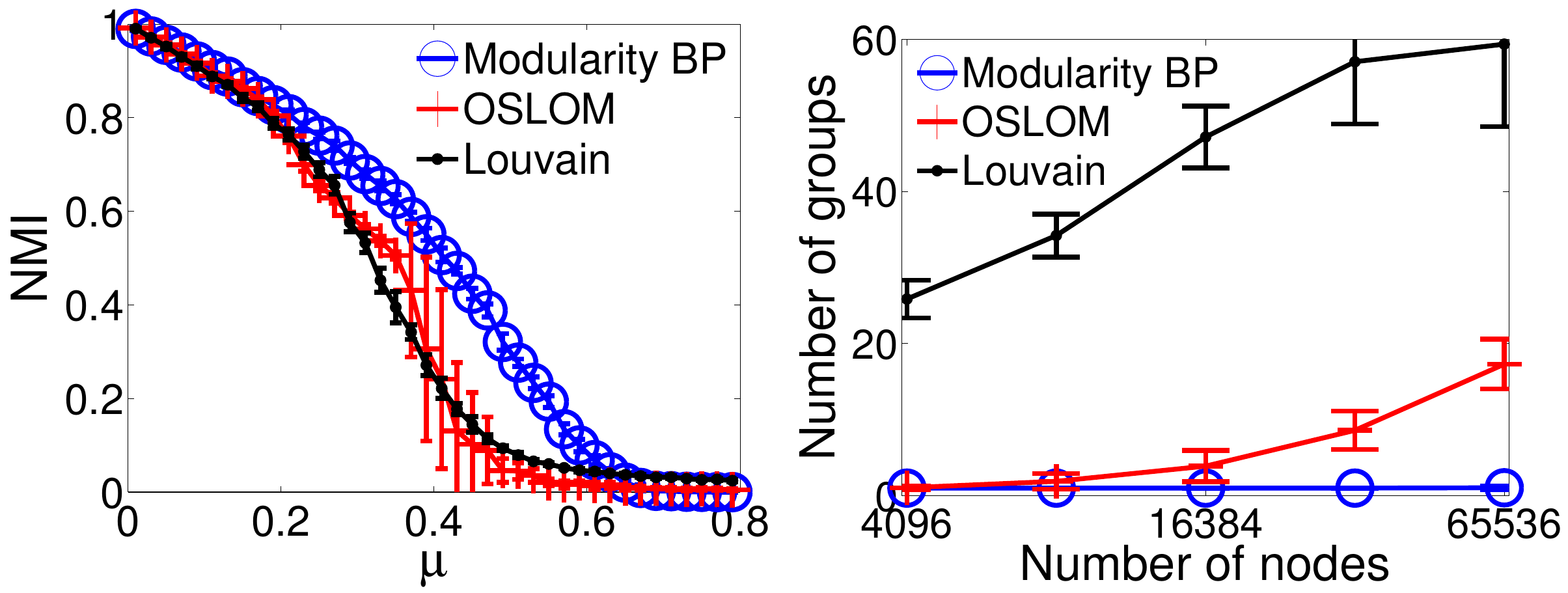} 
	\caption{Comparison of BP, the Louvain method, and OSLOM on benchmark networks with power-law degree distributions.  On the left, networks are LFR benchmarks with $n=10^4$ and $c=4$.  The distribution of community sizes follows a power law with exponent $-1$, ranging from $200$ to $400$.  The degree distribution is a power law with exponent $-2$, and the maximum degree is $30$.  We show the normalized mutual information (NMI) as a function of the mixing parameter $\mu$, and there is a range of $\mu$ where BP achieves a higher NMI than the other algorithms.  On the right, we show results on a random graph with no community structure, with a power law degree distribution with exponent $-2$ and mean $c=6$.  Here BP correctly chooses $q^*=1$ for the number of groups, while the other algorithms overfit, selecting a number of groups that grows with $n$. For both graphs, each data point is averaged over $20$ instances.  Compare Fig.~5 in the main text.
\label{fig:compare-lfr}
	}
\end{figure}

In Fig.~\ref{fig:compare-lfr} we show comparisons between our BP algorithm, Louvain~\cite{blondel2008fast}, and OSLOM~\cite{Lancichinetti-plosone} on networks with power-law degree distributions.  On the left, the graphs are generated by the LFR benchmark process~\cite{LFR-benchmark}.  We show the normalized mutual information~\cite{danon2005} as a function of the mixing parameter $\mu$.  As for the SBM graphs shown in the main text, there is a parameter range where BP achieves a higher NMI than the other algorithms.  On the right, we show results for a network with no community structure, where the degree distribution follows a power law with exponent $-2$.  While BP correctly chooses $q^*=1$ as the number of groups, the other algorithms overfit, finding a number of communities that grows with the network size.  These results are similar to those shown in Fig.~5 of the main text.

\section{The resolution limit}

In this section we describe results of our algorithm on the ring-of-cliques network, which is the standard example of the resolution limit~\cite{Fortunato2007}.  This network has size $n=ab$; it consists of $a$ cliques, each of which is composed of $b$ nodes, and which are connected to the neighboring cliques by a single link.  Thus the intuitively correct partition of the network puts each clique into one group.  However, when $b$ is sufficiently small compared to $a$, maximizing the modularity forces us to combine multiple cliques~\cite{Fortunato2007}.  For example, if $a=24$ and $b=5$, the correct partition with $24$ groups has modularity $0.8674$, while the division with $12$ groups of $2$ cliques each has modularity $0.8712$.  As a consequence, maximizing the modularity fails to divide the network correctly into the cliques.

\begin{figure*}
   \centering
\includegraphics[width=\fw\columnwidth]{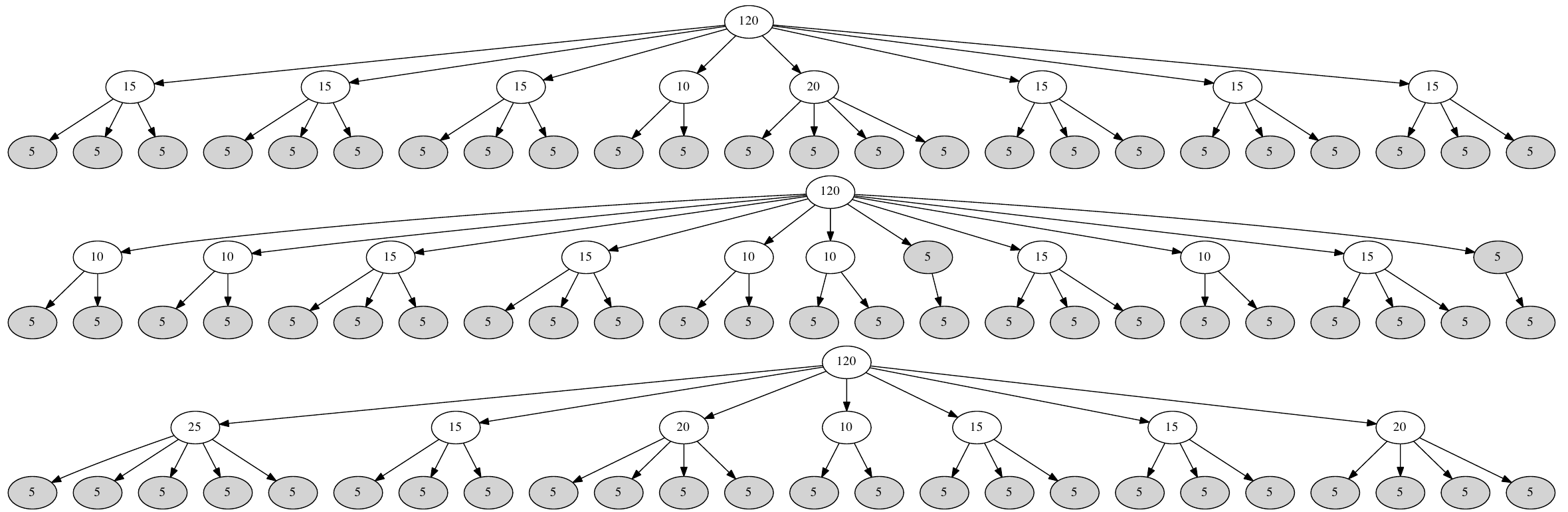} 
	\caption{Three dendrograms obtained by our hierarchical clustering algorithm on the ring of cliques, generated by independent runs with different initial conditions.  Here there are $a=24$ cliques of size $b=5$ each.  The number inside each node indicates the number of nodes in it.  In all three runs, the first level of splitting merges multiple cliques together, but the second level correctly divides the network into individual cliques.  This offers some evidence that our hierarchical algorithm can overcome the resolution limit, as opposed to algorithms that maximize the modularity. 
	\label{fig:dendo_ring}}
\end{figure*}

In Fig.~\ref{fig:dendo_ring} we plot the dendrogram obtained by our hierarchical clustering algorithm starting from $3$ different initial conditions (from top to bottom).  All three dendrograms have $2$ levels below the root.  The first split creates groups consisting of multiple cliques, but the second split correctly assigns each clique to its own group.  At that point the algorithm concludes that the cliques have no internal structure, and it stops subdividing.  This suggests that our hierarchical clustering algorithm may be able to avoid the resolution limit.

\end{appendix}

\end{document}